\documentclass[superscriptaddress,groupedaddress,nofootnoteinbib,11pt]{article}
\pdfoutput=1 
\usepackage{jheppub}
\usepackage{graphicx,color,amsmath}

\usepackage{graphicx}
\usepackage[caption=false]{subfig}
\usepackage{dcolumn}
\usepackage{multirow}
\usepackage{tabularx}
\usepackage{bm}
\usepackage{hyperref}
\usepackage{amsmath}
\usepackage{amssymb}
\usepackage{comment}
\usepackage{setspace}

\usepackage{color}
\usepackage{enumerate}
\usepackage{booktabs}
\usepackage{siunitx}
 \graphicspath{{figures/}}

\newcommand{\bea}{\begin{eqnarray}}
\newcommand{\eea}{\end{eqnarray}}

\definecolor{mypurple}{RGB}{164,64,214}

\begin{document}

\title{Searches for other vacua I: bubbles in our universe}


\author[a]{Anson Hook,}
\emailAdd{hook@umd.edu}
\author[b]{Junwu Huang}
\emailAdd{jhuang@perimeterinstitute.ca}

\affiliation[a]{Maryland Center for Fundamental Physics, University of Maryland, College Park, MD 20742, USA}

\affiliation[b]{Perimeter Institute for Theoretical Physics, Waterloo, Ontario N2L 2Y5, Canada}

\date{\today}

\abstract{We discuss models in which vacua other than our own can be directly observed in the present universe.  Models with density-dependent vacuum structure can give rise to `non-lethal'-vacua:  vacua with lower energy-density than our vacuum, but only in regions with finite Standard Model densities. These models provide an explicit example of a bubble which is confined to a finite region of space and produces potentially detectable signatures, unlike standard Coleman tunneling events where bubbles expand at the speed of light and are never directly observable. We study the expansion and contraction of a confined bubble created after a core-collapse supernova, focusing on energy deposition that may be observable in the vicinity of a supernova remnant due to the formation and evolution of a confined bubble.}

\maketitle

\section{Introduction}

The cosmological constant (CC) problem~\cite{Weinberg:1988cp} is the most intriguing problem in theoretical physics. Various dynamical or historical solutions~\cite{Abbott:1984qf,Brown:1987dd,Steinhardt:2006bf,Alberte:2016izw,Graham:2019bfu} have been proposed to address the CC problem. However, the only known viable solution to this problem is an anthropic one, proposed by Steven Weinberg in the 1970s~\cite{Weinberg:1987dv}. Weinberg's solution to the CC problem ``predicts" a small but non-zero cosmological constant, which was later confirmed by supernova and CMB measurements~\cite{Riess:1998cb,Perlmutter:1998np,Spergel:2003cb}. 

Weinberg's solution to the CC requires two major ingredients (see~\cite{Bousso:2007gp} for a review).  Firstly, it requires a landscape of vacua.  This landscape has many vacua so that in at least one of the them, just by pure chance, the cosmological constant is as small as the one we observe. Secondly, it calls for an anthropic principle, which postulates that only vacua which support galaxy formation from initial density perturbations can support life.  Since the proposal of this solution, both criteria have been studied extensively.  

There are two major obstructions to testing the existence of other vacua.  The first problem is that once a bubble of another vacuum has been created, it will rapidly either turn into a black hole or expand uncontrollably destroying our current vacuum and likely life as we know it.  The only observable signatures would be if there had been a bubble collision in the distant past and the collision had left an imprint in gravitational waves~\cite{Coleman:1980aw}.  The second problem is that the barrier between vacua may be extremely large resulting in the probability of creating a bubble of another vacuum being infinitesimally small~\cite{Coleman:1977py,Callan:1977pt}.

Both of these obstructions are present in many of the explicit constructions of a landscape of de Sitter vacua in string theory and field theory.  Perhaps the most notable attempt was that of Bousso and Polchinski~\cite{Bousso:2000xa,Polchinski:2006gy}. In their seminal paper in the 90s, they provide a construction of a string landscape where the distribution of vacua is mostly {\it random}: nearby vacua in field space in the landscape have cosmological constants that are maximally different from one another, and separated by barriers with height close to the string scale. These constructions led people to believe that direct tests of the string landscape are unlikely.

The first obstruction is the statement that the phase transition is likely to be disastrous if it happens now.  Inspiration around this problem can be drawn from a condensed matter system, e.g. a magnet.  We never worry that a magnetized piece of iron will eventually magnetize the whole universe.  The critical difference between the vacuum transition and the condensed matter system are finite density effects.  As shown in~\cite{Csaki:2018fls} and~\cite{Hook:2017psm}, new phases can exist in a very localized region inside compact objects, or a semi-local region ({\it confined bubbles}) if the landscape contains close by, degenerate or near degenerate vacua. These {\it confined bubbles} will be the main topic of this paper. 

As a field theory example of a confined bubble, take a vacuum with higher energy than our vacuum at zero density but with lower energy than our vacuum at finite density.  If a bubble of this vacuum is formed at finite density, then it will only expand a finite distance until the density becomes small enough that it is no longer energetically favorable for it to continue to expand.  Therefore, the bubble is confined to a non-trivial region of space, resulting in the ability to make non-lethal tests of the other vacua. Given that the highest density in our universe is $\sim (100 \,{\rm MeV})^4$ inside a neutron star while the CC is $\sim {\rm meV}^4$, the first obstruction of lethal bubbles can in fact be circumvented for $\sim (100 \,{\rm MeV})^4/{\rm meV}^4 \sim 10^{44}$ of the vacua in the landscape, leaving only the question of how reasonable is it to expect these bubbles to form in the first place.

Recently, several low energy constructions of the landscape suggest that, at least in an effective field theory (see~\cite{Graham:2015cka},~\cite{Arvanitaki:2016xds} and~\cite{Hook:2018jle}), it is possible to construct a landscape which is {\it ordered}: a situation where nearby vacua in the landscape have very similar cosmological constants, and have a low potential barrier between one another. In these constructions, some of the light fields can also have non-derivative couplings with the standard model matter, and even appear ``unnatural" due to the tuning of the cosmological constant~\cite{Arvanitaki:2016xds} or a discrete symmetry~\cite{Hook:2018jle}. In this paper, we will denote these light scalars or pseudo-scalars as {\it landscape fields}, and show a particular observational signature that is associated with the existence of many vacua in the potential of these landscape fields.

As discussed in~\cite{Arvanitaki:2016xds} and~\cite{Arvanitaki:2009fg}, one of the main requirements for a multiverse and anthropic solution to the electroweak hierarchy and cosmological constant problems is the existence of large volume extra dimensions and a zoo of light bosonic states that controls the sizes and shapes of these extra dimensions. These bosonic states, dubbed dilatons, moduli, string axions and dark photons, are the main target of low energy searches for beyond standard model physics. It has been suggested~\cite{Hu:2000ke,Hui:2016ltb} that light scalars as light as $10^{-22} \,{\rm eV}$ can be the dark matter of our universe, and can lead to interesting observable signatures when there are non-negligible self-couplings~\cite{FuzzyKen}. These new states, especially light scalars and pseudo-scalars that have significant self-couplings and non-derivative couplings with the standard model-and are therefore potentially unnatural-will usually have potentials that can be easily disturbed by the presence of matter. In fact, for a field as light as $m \sim 10^{-22} \,{\rm eV}$, assuming sub-Planckian field ranges and a technically natural potential, the height of potential can only be as large as $V \sim m^2 m_{\rm pl}^2 \sim (100 \,{\rm eV})^4$. This is much smaller than the densities in some of the densest objects in the universe, for example, a neutron star. If the scalar potential has period much smaller than the Planck scale (a scalar with attractive self-interactions), the potential barrier would be even shallower. 

The fact that the potentials of light states can be disturbed by a thermal population of particles they couple to has been known for a long time~\cite{Preskill:1982cy}. The QCD axion potential only appears when the temperature of the universe drops below $\Lambda_{\rm QCD}$; the Higgs potential receives thermal corrections at high temperature, which can potentially save the Higgs from instability after inflation~\cite{Espinosa:2007qp,Espinosa:2015qea,Franciolini:2018ebs} and lead to interesting observable signatures~\cite{NewMinimum2}. The effect of density~\cite{Hook:2017psm} was recently studied for a particular axion model and was shown to cause local phase transitions inside compact objects like a neutron star, which can be looked for with Advanced LIGO~\cite{Huang:2018pbu}. In some realizations of such an ``appears-to-be-tuned" axion~\cite{Hook:2018jle}, and a few other light scalars with``appears-to-be-unnatural" couplings, such a phase transition will result in the formation of a bubble.

In this paper, we discuss one particular example of a {\it confined bubble} which can give observable signatures near a supernova remnant. We only attempt to address the qualitative features of the bubble evolution, which can be achieved by analytical methods. In section~\ref{sec:setup}, we write a toy model of a potential and illustrate the features we will need from our landscape field, and how they qualitatively affect the bubble evolution. 
In section~\ref{Sec: formation}, we describe the formation of the bubble.
In sections~\ref{sec:expansion} and~\ref{sec:contraction}, we study the bubble expansion and contraction in the spherically symmetric limit. Section~\ref{sec:phenomonology} include some of the most striking signatures associated with an isolated bubble. In section~\ref{sec:collison}, we discuss some of the qualitative features of asymmetric bubbles near a neutron star as well as their phenomenological implications. We conclude in section~\ref{sec:conclusion} and discuss some of the future directions.

\section{A model that supports the creation of confined bubbles}\label{sec:setup}

In this section, we give an explicit example of a theory that supports both {\it confined bubbles} and their copious production at high densities while treating the background standard model densities as time independent. The effects associated with the dynamics of standard model matter in the presence of a bubble will be discussed in sections~\ref{sec:expansion} and~\ref{sec:contraction}.

We define a {\it confined bubble} as a region of the observable universe where a scalar field lives in a minimum that is different from the one we are in today in the solar system. A {\it confined bubble} and its production require the following conditions:
\begin{itemize}
  \item In empty space, there are many vacua that scan the cosmological constant~\footnote{To be precise, only two minima are needed for the existence of a {\it confined bubble}.  The requirement of many vacua is only present if the field also scans the cosmological constant.}
  \item In empty space, tunneling between vacua is highly suppressed
  \item ({\it Bubble Creation}) In medium, classical or quantum transitions between vacua are fast enough so that we can explore the other vacua of the landscape
  \item ({\it Localization}) The in medium vacua is localized to the areas of large density and do not take over and destroy our small CC universe 
\end{itemize}
Once the above-mentioned conditions are satisfied, a confined bubble will likely exist in our universe today.  
Confined bubbles appear in theories where there are degenerate or near-degenerate vacua in the landscape which can be significantly split by finite density effects.  This occurs whenever the scalar fields scanning the landscape has a coupling with the SM (see appendix~\ref{sec:example} for more details). In figure~\ref{fig:potential}, we show the qualitative features of a scalar potential that will allow a confined bubble. The potential has many minima in vacuum, guaranteeing an ability to scan the cosmological constant, while the dynamics and signatures we discuss in this paper mainly depend on two of the many minima: the standard model minimum $v_{\rm SM}$ and a different minimum $v_{\rm NEW}$.

\begin{figure}[ht]
\includegraphics[width=0.32\textwidth]{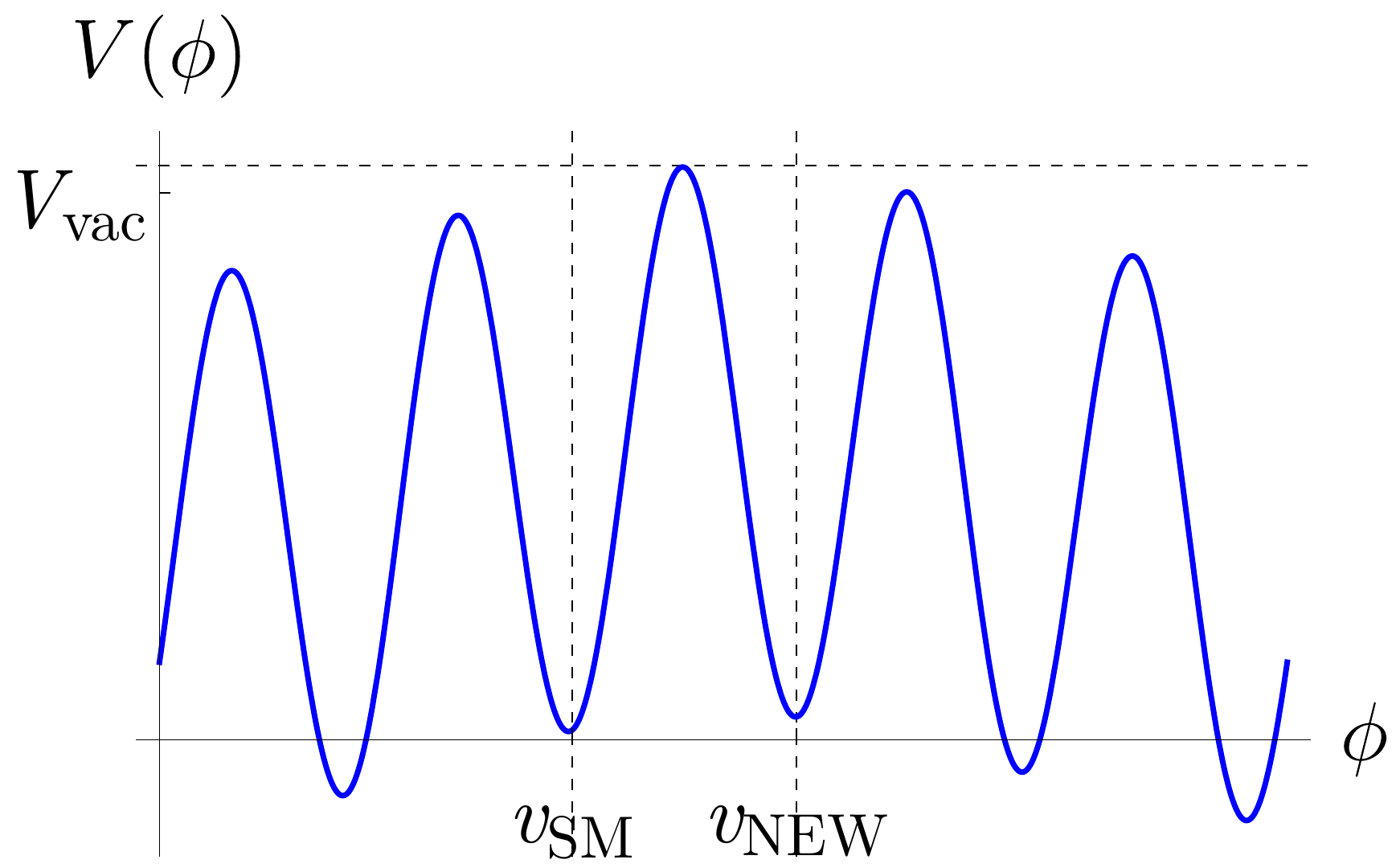}
\includegraphics[width=0.32\textwidth]{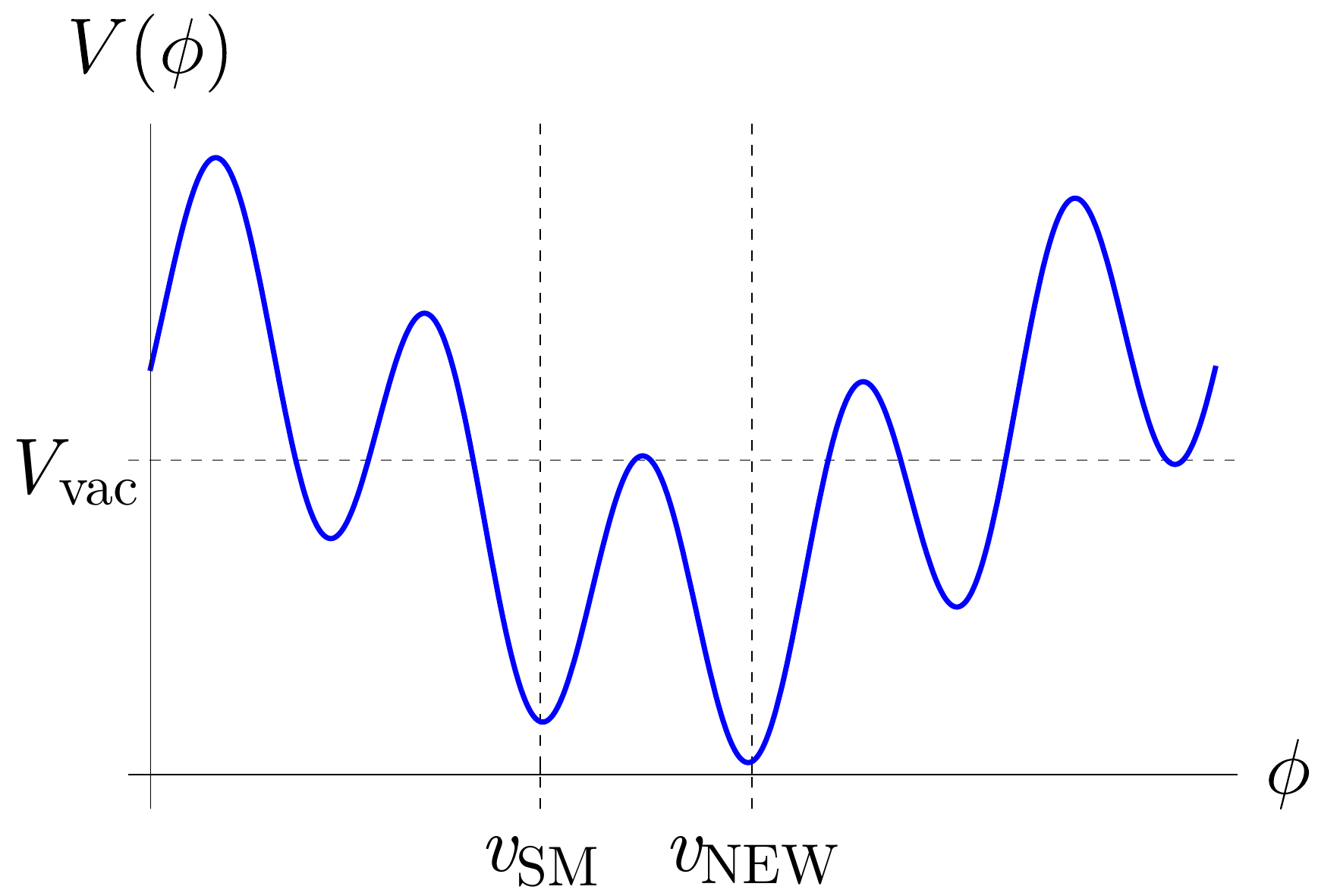}
\includegraphics[width=0.32\textwidth]{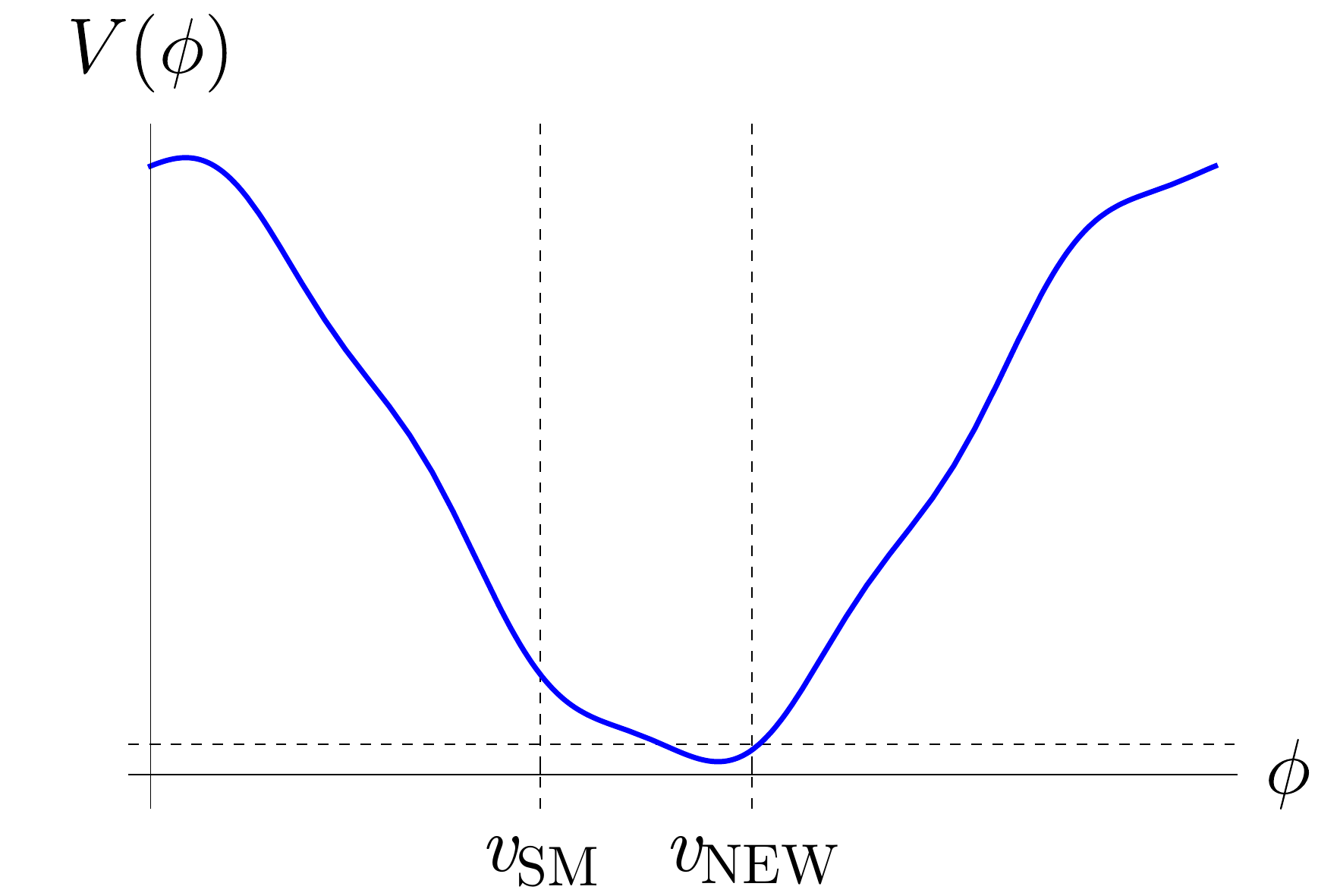}
\caption{A schematic picture of the form of a scalar potential in environments with different standard model densities that allow 
a confined bubble. (Left) At zero density, the standard model minimum $v_{\rm SM}$ and a different minimum $v_{\rm NEW}$ are nearly degenerate. (Middle) At intermediate density below the critical density, the standard model minimum $v_{\rm SM}$ gets lifted by scalar interactions with matter, and regions with $\phi = v_{\rm NEW}$ can expand. (Right) At high enough density, the barrier disappears and the standard model minimum is no longer a minimum of the theory, and regions with $\phi = v_{\rm NEW}$ can be classically generated. The horizontal dashed line shows the size of the barrier of the potential in empty space. In terms of the toy model in~\ref{sec:axionmodel}, the full field range shown in the figure is $\sim f_a$ while the field value difference between $v_{\rm SM}$ and $v_{\rm NEW}$ is $\sim f_a/N$.
}\label{fig:potential}
\end{figure}

\subsection{An explicit model}\label{sec:axionmodel}

The explicit example containing confined bubbles that we will consider is the case of a light axion~\cite{Hook:2018jle}. In this subsection, we provide a concise summary. A technically natural light axion has the potential
\begin{equation}
V \approx - \frac{m_a^2 f_a^2}{N^2} \cos \left(\frac{N a}{f_a}\right),\label{eq:Lag}
\end{equation}
where the axion $a$ is the {\it landscape field}, $f_a$ is the axion decay constant (setting the interaction strength between the axion field $a$ and the standard model matter) and $m_a^2 \approx m_{\pi}^2 f_{\pi}^2/ (2^N f_a^2)$ is the axion mass.
In empty space, the vacua of the potential are at $a = 2 \pi i f_a /N \, (i = 0 ,\,1,\,2\cdots)$ and are all degenerate (see~\cite{Hook:2017psm} for more detail). Such a degeneracy can be very weakly broken, e.g. by higher dimensional operators suppressed by the Planck scale~\cite{Kamionkowski:1992mf}.  As an explicit example, consider the case where the potential induced by these operators is $
V = - \epsilon^4 \cos \left ( a/F_a - \theta_0 \right )$
so that $a/F_a \approx \theta_0 $ is the true minimum, while all other minima of equation~\ref{eq:Lag} have slightly higher energy\footnote{We choose the minimum we currently live in to be $a=0$. $F_a$ is a scale that can be comparable to $f_a$ or much larger.}.  
These potentials combine to scan the cosmological constant without any observational effect in vacuum. For simplicity, we will first work in the limit where $\epsilon$ is small compared to everything else before moving on to larger values of $\epsilon$. That the potential can support confined bubbles is evident from the fact that the finite density contribution to the potential is~\cite{Hook:2017psm}
\bea
V_\text{finite} = (\sigma_N n_N)\cos \left(\frac{a}{f_a}\right)  \label{Eq:finitedensity}
\eea
where $\sigma_N = \sum_{q=u,d} m_q \frac{\partial m_N}{m_q} \approx 59\, {\rm MeV}$ is a coupling that can be found from lattice~\cite{Alarcon:2011zs} while $n_{N}$ is the number density of nucleons.  This finite density effect also breaks the degeneracy between vacua resulting in a true minimum near $i=N/2$.  In every other vacua, the vacuum energy is larger due to the corrections from nucleon density and the vacua becomes meta-stable.  
As a result, at finite density, $a/f_a \approx \pm \pi$ has lower energy than $a=0$ or the true vacuum in empty space with $a/F_a = \theta_0$.

The potential shown in Eq.~\ref{eq:Lag} not only supports confined bubbles but also provides a production mechanism.  In a very-high-density environments, such as a neutron star, the density might even be large enough that the minima $a = 2 \pi i f_a /N$ all disappear and only the minimum at $a/f_a \approx \pm\pi$ survives.  In this case, the previous vacua are all classically unstable and the field necessarily interpolates from $a = \pm \pi f_a$ around the high-density object to $a = 0$ far away.

In the following sections, we will be agnostic of the model that generates this {\it ordered} landscape of vacua. The generation, evolution, and observable signatures of the bubble can be described by the simple two-minima sub-structure of $v_{\rm SM}$ and $v_{\rm NEW}$ in the rich landscape shown in figure~\ref{fig:potential}.

\section{The formation of bubbles} \label{Sec: formation}

In this section, we discuss the formation of a bubble inside a region of very high density.
There are two possibilities for what occurs when the vacuum relaxes to the new in-medium minimum.  If the phase transition is first order, then the region containing a new phase is created by bubble nucleation and it will start to expand and speed up.  Eventually, it leaves the dense region of space and starts to slow down.  After a period of relaxation, it will finally arrive at a stationary situation.  
Alternatively, the transition to a new vacuum could be classical in nature.  In this case the entire finite density region might transition straight into the stationary solution rather than creating a bubble wall that expands and contracts. To determine what happens in this case, we numerically solved a toy model.  The toy model has a real scalar $\phi$ with the potential
\bea
V = -m^2 \phi^2/2 + \lambda \phi^4/4 + \epsilon \phi - \phi n_N.
\eea  
The small linear term $\epsilon$ biases the Mexican hat potential to favor the $\phi \approx -m/\sqrt{\lambda} = -v$ minimum.  $n_N$ is the density of particles and favors the $\phi \approx v$ minimum.  
We took a spherically symmetric expression for $n_N$ where it is a constant for a radius $r < R_c$ and falls off according to various power laws for $r > R_c$.  The density inside (for $r < R_c$) was chosen large enough so that the $\phi \approx - v$ minimum is no longer a minimum and is classically unstable towards rolling to the $\phi \approx v$ minimum.  Taking $\phi = -v$ as initial condition, we numerically solved for the evolution of the field $\phi$.  

We found that the entirety of the region $r < R_c$ started to roll together and a bubble wall is generated and ejected from the $r < R_c$ region.  This bubble expanded well beyond the equilibrium position for $\phi$ masses larger than the inverse size of the high-density region.  The requirement that the $\phi$ mass is larger than the inverse size of the high-density region is present so that gradient energy does not prevent the formation of a bubble altogether.  The final result is similar to what occurs in the case of bubble nucleation.  A bubble wall is generated and ejected well beyond the equilibrium position of the wall, before a combination of bubble wall tension and vacuum pressure cause it to start contracting.  This behavior occurs robustly and is independent of the parameters used as long as the density is large enough to destabilize the $\phi \approx -v$ minimum in the dense interior and large enough to accelerate the bubble into the less dense exterior.

As described before, during a process where the density of matter increases dramatically, in particular during a core collapse supernova, a bubble can be produced and launched outwards.  Back to our axion example, the axion field $a$ plays the role of the field $\phi$ in the numerical solution, while the neutron star radius $R_{\rm NS}$ and neutron number density play the role of $R_c$ and $n_N$, respectively, As the matter density increases beyond the critical density ($\sigma_N n_N \sim m_a^2 f_a^2/N^2$) such that the barrier disappears, the {\it landscape field} $a$ rolls down the scalar potential to the minimum $a = \pi f_a$ inside the dense region while the field is still at $a =0$ everywhere else. The oscillation of the scalar field has a period that is $\sim 1/m_a \ll R_{\rm NS}$ so that the bubble wall develops rapidly.  Because the bubble wall mass is much less than the total vacuum energy released, it is ejected at relativistic speeds.  Therefore, one can more or less study the dynamics of the scalar field assuming that the matter density profile is static.

Once a region where the scalar field is in the $a= \pi f_a$ minimum gets created inside the densest region of the supernova, such a region will generally want to expand into the surrounding less dense regions since the minimum where $a= \pi f_a$ has a lower energy compared to the minimum at $a=0$. During this step and subsequent steps, we can approximate the field profile as a thin bubble wall since $1/m_a$ is much smaller than the size of the neutron star in most of the interesting parameter space\footnote{In the discussions of the bubble expansion in section~\ref{sec:expansion}, we will start the expansion assuming a non-zero initial velocity for the field profile. However, this is not necessary as the field profile generally gains speed through interactions with hot matter inside and outside of the bubble. Additionally, as we will discuss in more details in section~\ref{sec:shock}, a shock wave of standard model matter can usually help push the bubble out.}. As is familiar from the evolution of Coleman thin wall bubbles in empty space, as the bubble wall expands, the energy difference between the two minima in matter is transferred into the kinetic energy of the bubble wall.  As the local density of matter decreases while the bubble moves out, the energy differences between the two minima decreases and eventually is not sufficient to support the bubble expansion against brane tension, and the bubble wall slows down and comes to a stop. During the whole process, a significant portion of the mass energy of the standard model matter inside the progenitor gets stored inside the bubble wall as particles drop inside the bubble, and the bubble wall can carry a mass that is $\mathcal{O}(1 \%)$ of the total mass of the progenitor. Section~\ref{sec:expansion} contains a more detailed discussion.

The bubble wall expansion will eventually come to a stop due to brane tension. Quantitative understanding of the contraction phase depends strongly on treating the standard model plasma properly and cannot be achieved with the simple approximation used in this section. Section~\ref{sec:contraction} contains a more detailed discussion.

\section{Bubble expansion}\label{sec:expansion}

In this section, we study the bubble expansion taking into account the standard model plasma and its interactions with the expanding bubble. At the same time, we will remain agnostic about the details for the interaction between the standard model matter and the bubble wall but treat it as a change of the mass of the different states inside and outside of the bubble wall.  These calculations are completely analogous to those done in the context of expanding bubble walls during the Electroweak phase transition in the early universe~\cite{Linde:1981zj,Linde:1980tt,Dine:1992wr}, with the important caveat that a relativistic plasma has a much larger pressure than a non-relativistic plasma.

The initial acceleration of the bubble is provided by the pressure of the particles inside the bubble as well as the pressure that results from particles outside getting accelerated as they cross the bubble wall. For example, in the particular case of a bubble wall constructed out of the QCD axion, as baryons outside the bubble ($\theta=0$) penetrate the bubble wall to get inside the bubble ($\theta=\pi$), the baryon mass $m$ decreases by $\delta m \sim \sigma_N$ (see definition of $\sigma_N$ below equation~\ref{Eq:finitedensity}). Such a mass decrease as particles drop into the region in the new minimum can also be seen from the fact that the energy of the standard model minimum is lifted compared to that of the new minimum in high-density matter in figure~\ref{fig:potential}. During this process, the baryons might penetrate the bubble wall and accelerate, or get reflected by the bubble wall. The transmission and reflection coefficient can be calculated using 1D Quantum Mechanics in the limit where the particles start at rest:

\begin{eqnarray}
\mathbb{T} &= \frac{4 k' k}{(k+k')^2}, \nonumber\\
\mathbb{R} &= \frac{(k'-k)^2}{(k+k')^2},
\end{eqnarray}
where $k$ and $k'$ are the momentum of the incident and transmitted wave in the reference frame of the bubble wall and $\mathbb{T}$ and $\mathbb{R}$ are the transmission and reflection coefficients. The momentum the bubble gets from each particle is 

\begin{equation}
\delta p_{\rm bubble} = (k'-k)\mathbb{T} - 2k \mathbb{R} = 2k\left(\frac{k'-k}{k'+k}\right),\label{eq:dpnr}
\end{equation}
pointing towards the outside of the bubble. Therefore, the average effect of every particle that starts off outside of the bubble is to accelerate the bubble.

For a non-relativistic bubble, $ k = m v_{\rm wall}$ while $ k' = \sqrt{k^2 + 2 m \delta m}$. If the wall velocity $v_{\rm wall} \ll \left(\frac{\delta m}{m}\right)^{1/2}$, the momentum that the transmitted particles get is much larger than the incident momentum ($k'\gg k$). In this case, the transmission coefficient $\mathbb{T} \sim k/k' \ll 1$ while the pressure on the bubble is approximately
\begin{equation}
P = n_B v_{\rm wall} \delta p_{\rm bubble} = 2 \rho_B v_{\rm wall}^2,
\end{equation}
where $n_B$ and $\rho_B$ are the number and energy density of baryonic matter near the compact neutron star.  The pressure is still positive despite the large reflection coefficient because whenever a rare particle is transmitted, it transfers a large momentum kick to the wall.

For a relativistic bubble, $ k = \gamma_{\rm wall} m v_{\rm wall}$ and $k'-k \approx \delta m/\gamma_{\rm wall} v_{\rm wall} \ll k$. In this case, most of the baryons the bubble encounters in its expansion will penetrate the bubble wall. The pressure on the bubble is approximately
\begin{equation}
P = n_B \gamma_{\rm wall} v_{\rm wall} \delta p_{\rm bubble} = \rho_B \left(\frac{\delta m}{m}\right).
\end{equation}

In summary, we can find the equation of motion of the bubble wall (in the non-relativistic and ultra-relativistic limit) due to its interactions with matter as
\begin{align}
 \begin{cases}
\frac{{\rm d} \gamma_{\rm wall}}{{\rm d} R} = \frac{n_B \delta m}{\mathcal{T}}-2\frac{\gamma_{\rm wall}^2 +\gamma_{\rm wall}-1}{\gamma_{\rm wall} R}, & {\rm for} \, v_{\rm wall} \gg \left(\frac{\delta m}{m}\right)^{1/2}  \\
\frac{{\rm d} v_{\rm wall}}{{\rm d} t} = \frac{2 n_B m v_{\rm wall}^2}{\mathcal{T}}-\frac{2}{R}, & {\rm for}\, v_{\rm wall} \ll \left(\frac{\delta m}{m}\right)^{1/2} 
\end{cases},\label{eq:expansion}
\end{align}
where $\mathcal{T}$ is the brane tension and $R$ is the radius of the bubble, and the velocity $v_{\rm wall}$ should be replaced by $v_{\rm gas}$ in the case where the standard model plasma collective motion is comparable or faster than that of the bubble wall.  The first term in the differential equation is the pressure due to matter interacting with the wall.  The second term is due to tension and the increase in the mass of the wall as it expands.  To simplify matters, we assume, for now, that the difference in vacuum energy is small enough to be irrelevant to the dynamics of the bubble.

Using Eq.~\ref{eq:expansion}, we find that after its initial launch by the supernova shock, the bubble will get accelerated by its interactions with the gas outside the bubble to ultra-relativistic speed very close to the surface of the neutron star. It reaches a maximal boost factor
\begin{equation}
\gamma_{\rm max} \approx \int_{R_{\rm NS}}^{R_{\rm max}} \frac{\delta m}{m} \frac{\rho_B (R)}{\mathcal{T}} \mathrm{d} R,
\end{equation}
where $R_{\rm max}$ is the maximal size of the bubble. For a density profile outside of a neutron star during a supernova, $\rho_B (R)$ decreases as roughly $\sim 1/R^3$ outside the neutron star. Therefore, the integral is dominated at short distances close to the neutron star, and can be evaluated to be approximately
\begin{equation}
\gamma_{\rm max} \approx \frac{\delta m n_{\rm NS} R_{\rm NS}}{\mathcal{T}},
\end{equation}
reached at $R \sim 2 R_{\rm NS}$.

After the initial large boost, the bubble will continue to expand into the surrounding matter, slowly converting its kinetic energy into potential energy stored in brane tension while getting accelerated by the baryons entering the bubble. Such an expansion stops when roughly all the energy is stored in the tension of the brane.
\begin{equation}
\gamma_{\rm max} R_{\rm NS}^2 = R_{\rm max}^2.
\end{equation}
Because the initial boost is not particularly sensitive to the detailed density distribution, the maximal size reached by the bubble is roughly always the same (up to a logarithmic factor):
\begin{equation}
R_{\rm max} = R_{\rm NS} \left(\frac{\delta m n_{\rm NS} R_{\rm NS}}{\mathcal{T}}\right)^{1/2}.
\end{equation}
Such a maximal radius is parametrically the same as the radius where a dynamical equilibrium is reached between the gas pressure and the brane tension for an $1/R^3$ density fall off $n(R_{\rm max}) \delta m R_{\rm max} \approx \mathcal{T}$ in the effective potential picture. 

The expansion phase is also affected by the energy differences between the two minima in empty space. If $\Delta V = V(v_{\rm SM}) - V(v_{\rm NEW})>0$, the minimum we are in is the false vacuum and the bubble expansion is assisted by this energy difference.  For $\Delta V > \mathcal{T}/R_{\rm max}$ and $\Delta V >0$, the bubble deceleration will stop when $R= \mathcal{T}/\Delta V$ and will instead start accelerating and expand indefinitely.  The bubble will expand past its critical radius and proceed to take over the whole observable universe.  Such a bubble will encounter us at the speed of light and does not satisfy the {\it localization} criterion. 

On the other hand, if $\Delta V <0$, the minimum we are in right now is the true vacuum and the energy difference will help slow down the bubble in its expansion. For $|\Delta V| > \mathcal{T}/R_{\rm max}$, the bubble does not reach $R_{\rm max}$ in its expansion. Rather, it stops when
\begin{equation}
4\pi \gamma_{\rm max} \mathcal{T} R_{\rm NS}^2 \approx\frac{4\pi}{3}|\Delta V| \tilde{R}_{\rm max}^3 \rightarrow \tilde{R}_{\rm max} = \left(\frac{3 \gamma_{\rm max} \mathcal{T} R_{\rm NS}^2}{\left|\Delta V\right|}\right)^{1/3}.
\end{equation}
In the case where $\frac{\mathcal{T}}{|\Delta V|}$ is much larger than the size of the neutron star, adding the vacuum energy changes the final radius of the bubble.  In the other limit, the bubble containing the neutron star will never grow outside the neutron star and remain confined to its surface.  The signatures will be similar to those discussed in Ref.~\cite{Hook:2017psm}.

\begin{figure}[ht]
\includegraphics[scale=0.5]{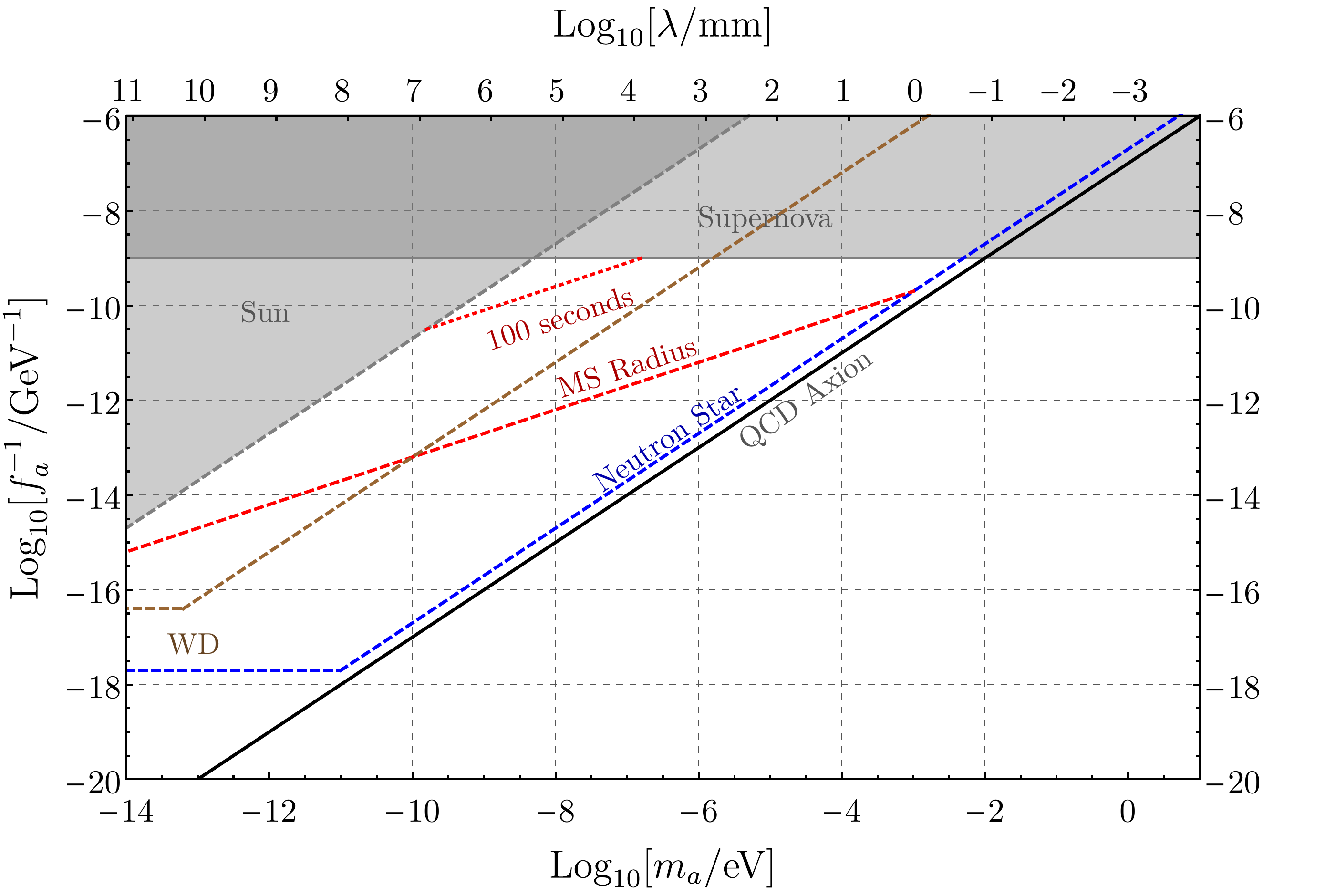}
\caption{Parameter space of light axions with matter couplings shown in section~\ref{sec:axionmodel}.  Supernova constraints~\cite{Chang:2018rso} and constraints coming from the measurement of solar neutrinos~\cite{Hook:2017psm} are shown in shaded grey.  Bubbles can be launched by a neutron star in the region to the left of the neutron star line (blue dashed).  The bubbles will expand to roughly the size of a main sequence star at the lower red dashed line, and to a radius of $100 \,{\rm seconds}$ at the upper red dashed line.
}\label{fig:axion}
\end{figure}

\begin{figure}[ht]
\includegraphics[scale=0.5]{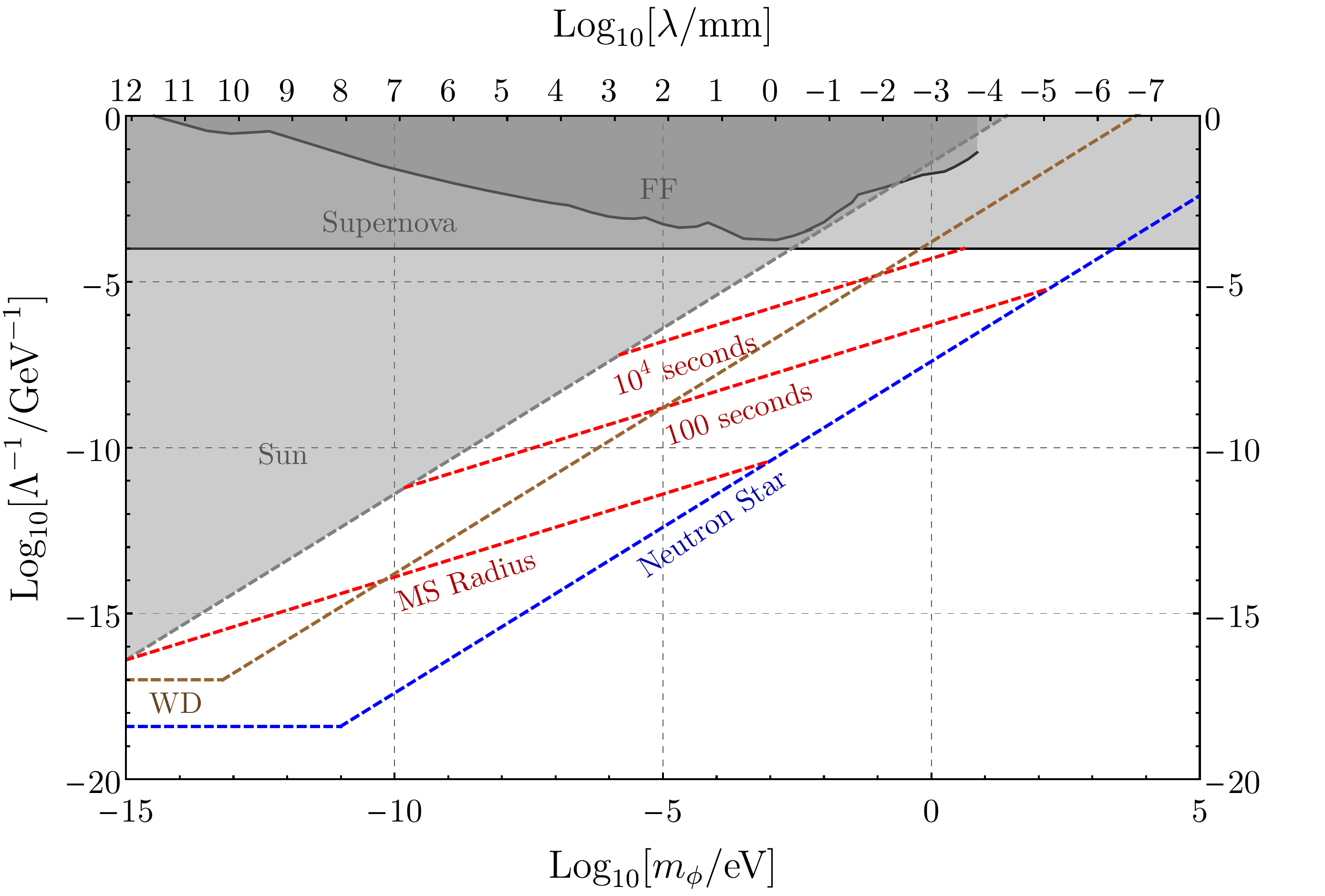}
\caption{Parameter space of scalars which interact with nucleons as $\frac{\phi^2}{\Lambda^2} \mathcal{L}_{\rm SM}$. The shaded regions are excluded by supernova~\cite{Chang:2018rso}, measurements of solar neutrinos~\cite{Hook:2017psm}, and fifth-force measurements (see~\cite{Arvanitaki:2014faa} and references within). Bubbles can be launched by a neutron star in the region to the left of the neutron star line (blue dashed). The bubbles will expand to a size with roughly main sequence star radius above the lower red dashed line, and 100 seconds ($10^4$ seconds) radius above the middle (upper) red dashed line. 
}\label{fig:dg}
\end{figure}

To conclude, in this section, we studied the expansion of a bubble following bubble generation during a core collapse supernova. In figure \ref{fig:axion} and \ref{fig:dg}, we summarized the sizes a bubble can reach ($R_{\rm max}$) for two different types of interactions. Bubble can reach sizes as large as hundreds of light seconds, many orders of magnitude larger than the size of the forming neutron star ($R_{\rm NS}\sim 10\,{\rm km}$). In the plots, we also show current constraints coming from supernova cooling, fifth force experiments, and the requirement that the Sun is not dense enough to source a bubble (see~\cite{Chang:2018rso},~\cite{Hook:2017psm} and  references within). 

\section{Bubble contraction}\label{sec:contraction}

In this section, we will discuss the contraction phase of the bubble that occurs after it has reached its maximal size due to its brane tension. The contraction phase depends only quantitatively on the difference of energy of the two vacua. Similar to the expansion phase, we will focus on the case where $\Delta V \approx 0$ and comment on the case where $\Delta V \neq 0$ when it becomes important.

For presentation purposes, we will first take the approximation that the initial shock wave has already passed through the bubble and the density within the bubble can be approximated by a power law of $1/R^{\alpha}$ and leave the discussions of the shock to a separate subsection. The density around the supernova shortly after the supernova passes through can be best fitted by $\alpha \approx 3$~\cite{1986AA}, which we will use for most of the discussion in this section.  The general features of the observable signatures are independent of the exact density profile around the supernova, while the magnitude is sensitive to the exact density profile.  We leave the detailed calculations of the size of the effects to future work.

There are two types of contraction phases.  The first phase is a fast contraction phase.  Equation~\ref{eq:expansion} respects a $t \rightarrow -t$ symmetry so that because there exists a fast expansion solution, there also exists a fast contraction solution.  As will be discussed in more detail in appendix~\ref{Sec: fastosc}, this fast contraction phase, if ever present, lasts for only a short time before friction ends this fast oscillation phase after a few cycles, making it mostly phenomenologically irrelevant. The second phase is a slow contraction phase and is the main subject of this section.

\begin{figure}[ht]
\centering
\includegraphics[scale=0.42]{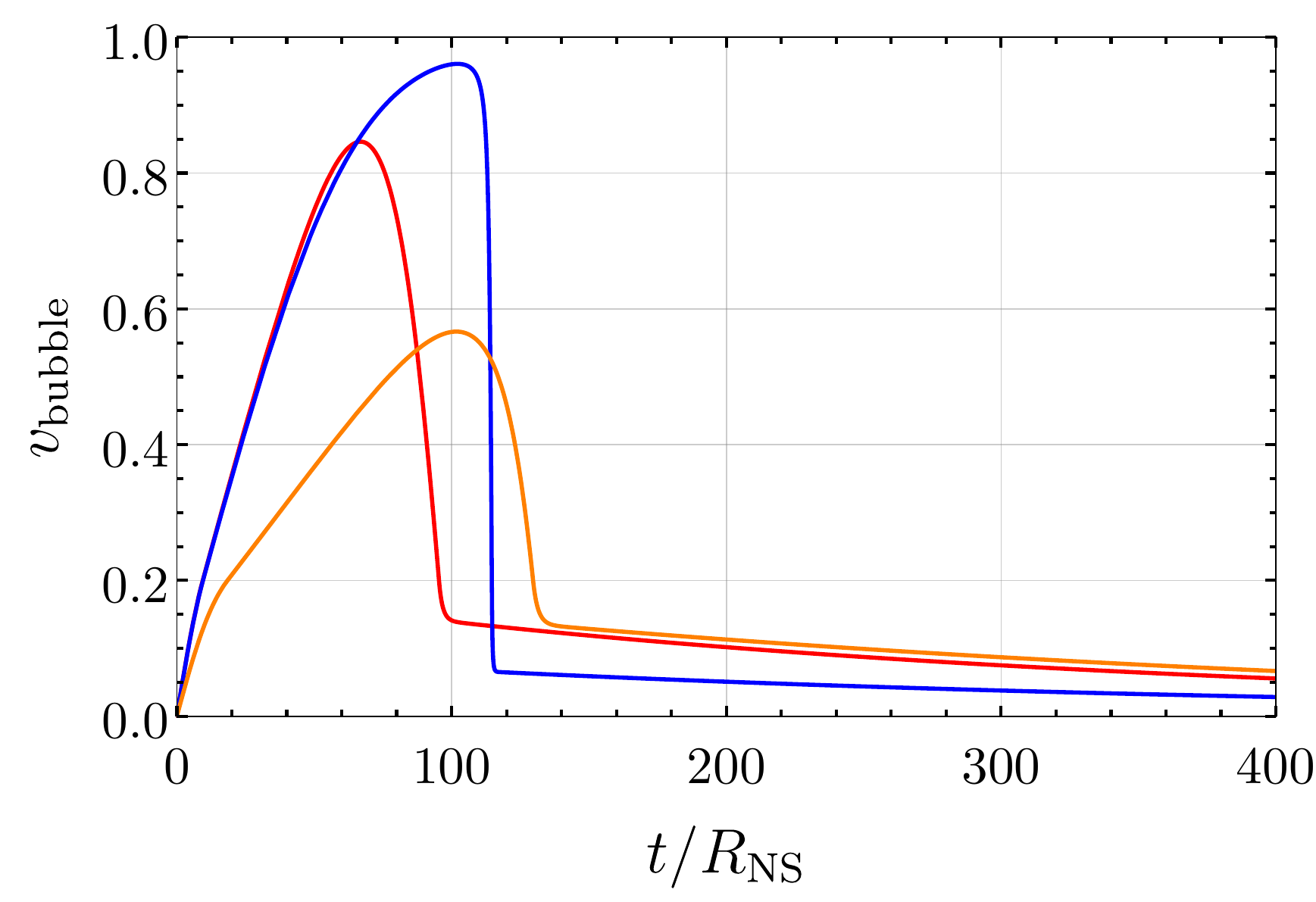}
\includegraphics[scale=0.42]{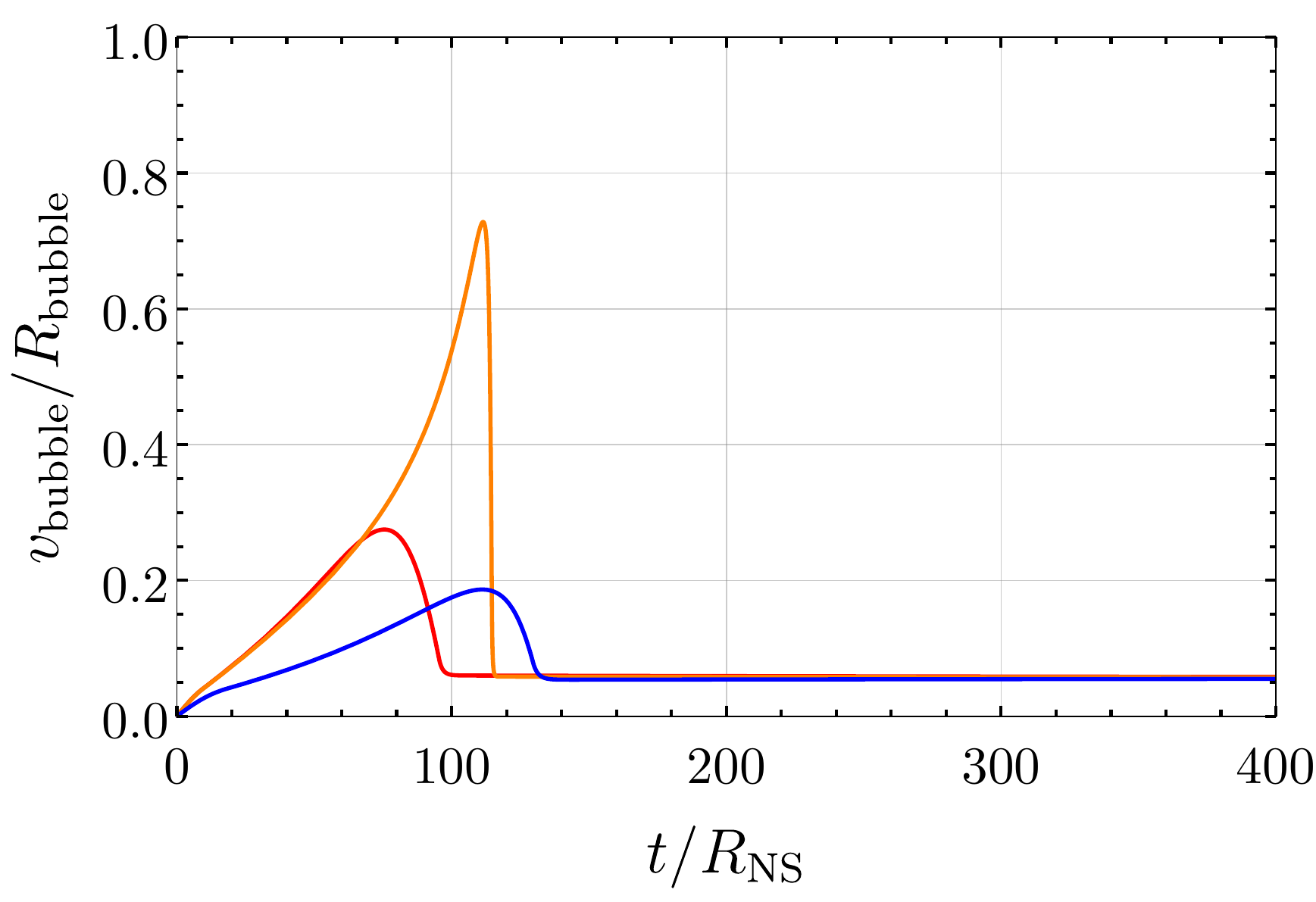}
\caption{The evolution of a contracting bubble with different size of energy differences between the two vacua $\Delta V$. The blue, red and orange lines correspond to cases where $\Delta V R_{\rm max}/T  = \{-1/2,0,1/2\}$, respectively. The bubbles in these plots have a $R_{\rm max} = 100 R_{\rm NS}$ when $\Delta V =0$. The left figure shows how the bubble velocity evolves as a function of time with $(\delta m/m)^{1/2} =0.2$. The right figure highlights the late time evolution where $v/R$ (arbitrary unit) approaches a constant.}\label{fig:contraction}
\end{figure}

During the initial stage of bubble contraction, the bubble will accelerate into very dilute matter due to brane tension $\mathcal{T}$ and the energy difference between the two minima $\Delta V$.  While the initial rate of acceleration depends strongly on the relative size of the brane tension and the difference in vacuum energy, the bubble quickly reaches a terminal velocity that is only weakly dependent on either $\mathcal{T}$ or $\Delta V$ (see figure~\ref{fig:contraction}). The mildly relativistic terminal velocity, usually between $0.3 \, c$ and $0.9 \, c$, is reached due to the balance between $\mathcal{T}$ accelerating the bubble towards smaller radii and the friction force due to bubble-matter interactions, which grows as the bubble shrinks. 

Similar to the expansion phase, one of the most important parameters governing the slow contraction phase is the critical velocity $(\delta m/m)^{1/2}$. When the bubble is contracting faster than $(\delta m/m)^{1/2}$, the friction due to matter is independent of the velocity of the bubble.  When the bubble is contracting slower than than $(\delta m/m)^{1/2}$, the deceleration becomes proportional to $v_{\rm wall}^2$. During this phase, the bubble size shrinks by an $\mathcal{O}(1)$ amount during a time period that is approximately $\sim R_{\rm max}$.
 
Once the bubble wall becomes slower than $(\delta m/m)^{1/2}$, it enters the second stage of the slow contraction phase.  It is roughly described by a damped oscillator, where the velocity of the bubble decreases slowly as the bubble shrinks. In this phase, the brane tension $\mathcal{T}$ (and the energy difference between the two minima in empty space $\Delta V$) are balanced by the friction force due to interactions with matter. In this stage, the dynamics of the bubble can be described by the simple equation\footnote{In this subsection, we do not take into account the accumulation of matter as the bubble slowly contracts. Such an accumulation depends on the density transport and heat conductance in the surroundings of the neutron star, which we do not know how to treat properly without detailed numerical simulation. In reality, such an accumulation effect will change how the density increases as the bubble contracts. For a density profile that is $ n(R) \propto R^{-3}$, this is likely a logarithmic increase.}:
\begin{equation}
\frac{\mathrm{d}^2 R}{\mathrm{d} t^2} = \frac{n(R) m v^2}{2 \mathcal{T}} -\frac{2}{R}- \frac{\Delta V}{\mathcal{T}}.
\end{equation}
For $ n(R) \propto R^{-3}$, this results in a scaling solution where the bubble velocity
\begin{equation}
v_{\rm b} \propto R \left(1 + \frac{\Delta V R}{\mathcal{T}}\right)^{1/2}.
\end{equation}

The second stage of contraction comes to an end when the bubble velocity decreases to be comparable to the velocity of the standard model plasma confined inside the bubble, in which case the friction force starts to grow as $R^{-3}$ as the bubble shrinks. In this last stage, the bubble velocity will quickly decrease to close to zero.
The contraction due to $\mathcal{T}$ and $\Delta V$ is counteracted by a constant thermal pressure of the material inside.
As the thermal matter inside slowly radiates away its energy in photons, the bubble slowly contracts.  Such a dynamical equilibrium is described by
\begin{eqnarray}
\frac{\mathrm{d} R}{\mathrm{d} t} &=& \frac{\sigma T^4 R}{2 \mathcal{T} + \Delta V R},\nonumber\\
\frac{n(R) T}{\mathcal{T}} &=& \frac{2}{R} + \frac{\Delta V}{\mathcal{T}},\label{eq:phase3evo}
\end{eqnarray}
where $\sigma$ is the Stefan-Boltzmann constant. In equation~\ref{eq:phase3evo}, we assume that the standard model matter density outside the bubble wall is negligible while the density inside of the bubble is $n(R)$, since Standard Model matter does not have enough kinetic energy to exit the bubble during this slow contraction phase. We assume the emission from the region inside the bubble is mainly from the surface during this stage due to the fact that the bubble usually has already shrunk enough such that it is already in regions where the density of matter is high enough such that it is opaque to the radiation from inside. We assume the emission from the bubble has a blackbody spectrum because such a phase is extremely long such that the standard model matter inside the bubble always has time to thermalize. Given a density profile of standard model matter $n(R)$, we can solve for temperature $T$ and radius $R$, and as a result thermal emission luminosity $L$, as a function of time. The first part of equation~\ref{eq:phase3evo} describes energy conservation, while the second is basically the Young-Laplace equation in the spherically symmetric limit. For a density that decreases as $1/R^3$, this results in a dependence
\begin{equation}
T \propto  t^{-1/4}, \quad R \propto  t^{-1/8}, \quad L \propto  t^{-5/4},\label{eq:phase3scaling}
\end{equation}
for a bubble contracting due to brane tension. Note that the matter inside the bubble cools a lot slower than normal thermal emission because the bubble wall is injecting its rest mass as energy into the system.
As a result, the anomalous thermal emission will be at a relatively high temperature and luminosity long after the supernova has taken place (see figure~\ref{fig:temp} and~\ref{fig:power}). This also implies that the final long term observational signature is largely independent of the earlier stage of the bubble evolution, including the dynamics of the shock wave, bubble irregularities and matter accumulation.

The bubble evolution will eventually come to an end when the bubble slowly lands on the neutron star. In most of the previous discussion, we have not taken into account the temperature evolution of the neutron star itself. We will postpone these discussions to section~\ref{sec:glow}.

\subsection{Shock wave}\label{sec:shock}

One particularly interesting element of bubble evolution after a supernova explosion is the expansion of the shock wave (see~\cite{RevModPhys.62.801} and references within). A core collapse supernova starts when the compact inert core of a massive star exceeds the Chandrasekhar limit, where the electron degeneracy pressure ceases to be large enough to counter the gravitational pressure. The inner core will undergo a rapid implosion phase while the outer core, losing its support from the inner core, also collapses due to gravity. The imploding material accelerates to relativistic speed before bouncing off of the inner core, whose collapse is terminated by neutron degeneracy pressure. The now outgoing shock wave loses energy as it penetrates the material surrounding the proto-neutron star and comes to a stall as it photo-dissociates the surrounding material.  The shock is assumed to be revived when the neutrinos produced in the supernova core heat up the materials around the shock and create a temperature and pressure gradient near the shock location, which re-accelerates the shock wave. The shock wave is accelerated to very high velocities close to the speed of light as it eventually expands out of the progenitor star. 

\paragraph{Initial expansion}: The shock wave can be important for the initial expansion of the bubble. Even if the field profile does not expand out of a neutron star fast enough due to a lack of initial velocity, the initially fast expanding shock wave will have enough kinetic energy to push out the scalar field bubble through its interaction with the bubble wall (see equation~\ref{eq:expansion} with $v_{\rm wall}$ replaced by the relative velocity between the shock and the bubble wall). The bubble wall and the shock wave quickly decouples as the shock wave comes to a stall while the bubble wall continues to accelerate to relativistic speed during its expansion.  A full 3D simulation is likely needed to understand the complete effect of the shock wave on the bubble as well as how much the expanding bubble affects the supernova explosion. However, the details of shock wave dynamics should not strongly affect the signature we propose to look for.

\paragraph{Subsequent contraction}: The interaction between the shock wave and a contracting bubble wall could potentially alter the supernova enough to lead to constraints.  Though a qualitative picture of a core collapse supernova has been understood for decades, there are still a lot of uncertainties regarding the time-dependent density and temperature profile of the surroundings of the proto-neutron star as the shock wave is propagating out. As a result, there is an order-of-magnitude uncertainty on the velocity of the shock wave during its initial expansion~\cite{1976ApJS}.

The properties of the shock wave can be significantly affected by the presence of a contracting bubble. As an expanding shock wave with velocity $v_{\rm ini}$ hits a contracting bubble wave with velocity $v_{\rm wall}$, the shock wave slows down by
\begin{equation}
v_{\rm fi} = v_{\rm ini} - 2 \left(\frac{ \delta m }{ m}\right)\left(\frac{1+v_{\rm ini} v_{\rm wall}}{v_{\rm ini}+v_{\rm wall}}\right),
\end{equation}
in the limit where $ \frac{\delta m }{m} \ll 1$ and the shock wave is not ultra-relativistic. In the limit where the wall is not moving ($v_{\rm wall}=0$), this requirement simplifies to that the particles inside the shock wave need to carry enough kinetic energy to exit the bubble. 
As the velocity of the bubble increases, the reflection coefficient goes down and it gets easier for the shock wave to pass through the bubble. However even in the relativistic limit, the final velocity of the shock wave can still point inwards if the initial velocity is not large enough.  

Requiring that the shock wave exits the bubble and continue its outward expansion would constrain the differences in the mass of particles inside versus outside of the bubble. Unfortunately, there is a huge uncertainty in the predictions of the maximal velocity of the shock wave after the shock is revived, ranging from $10^{-2}\, c$ to $0.3 \, c$, as well as the time dependence of the shock wave velocity~\cite{2015MNRAS.453..287M}. Therefore, the measurement of the supernova shock wave at late times does not place a strong constraint on the size of $\delta m/m$; a $\delta m/m$ larger than $0.1$ may be in tension with the observation of various supernova remnants, however it is difficult to be certain. With improved simulations of the supernova shock wave expansion, in particular a simulation of the supernova environment with the additional scalar field, one might place a much stronger bound on $\delta m/m$ of a proton or a neutron. These simulations would have strong implications for particles with QCD axion like couplings, since the fractional mass shift caused by these couplings will necessarily be $\delta m/m \approx \sigma_N/m_N \approx 6\,\%$.

\section{Phenomenology of isolated spherical bubbles}\label{sec:phenomonology}

\subsection{A glowing bubble}\label{sec:glow}

The most important and potentially long-lasting phenomenological signature of the confined bubble comes from the contraction phase of the bubble. As discussed in section~\ref{sec:contraction}, in the last stage of the bubble evolution, the bubble releases its kinetic and mass energy into the thermal energy of the plasma inside the bubble, which gets radiated away by thermal or non-thermal radiation.  For bubbles that are relatively small in size, in particular if there is no significant separation between the maximal bubble size and the size of the neutron star, the bubble evolution happens for a relatively short period of time, and we cannot neglect the presence of a much hotter neutron star and its effect on the densities between the bubble and the neutron star~\cite{Yakovlev:2004iq}. 

However, for bubbles that can potentially grow to sizes of order light seconds and beyond (see figures~\ref{fig:axion} and~\ref{fig:dg}), it is conceivable that dynamics are dominated by the matter near the bubble rather than the neutron star at the origin.  These densities can come from the part of the outgoing shock wave that gets reflected by the bubble wall (see section~\ref{sec:shock}), as well as the residue density that is not carried away by the shock wave during the shock expansion. In the case of a binary system, the matter density can also come from the companion which donates matter to the neutron star.  For bubbles with smaller brane tensions, which are also the ones that grow to the largest sizes, the density required to support a bubble can be as small as $10^{-6} {\rm g/cm^3}$, much smaller than the density of the core or the progenitor.

\begin{figure}[t]
\centering
\includegraphics[scale=0.4]{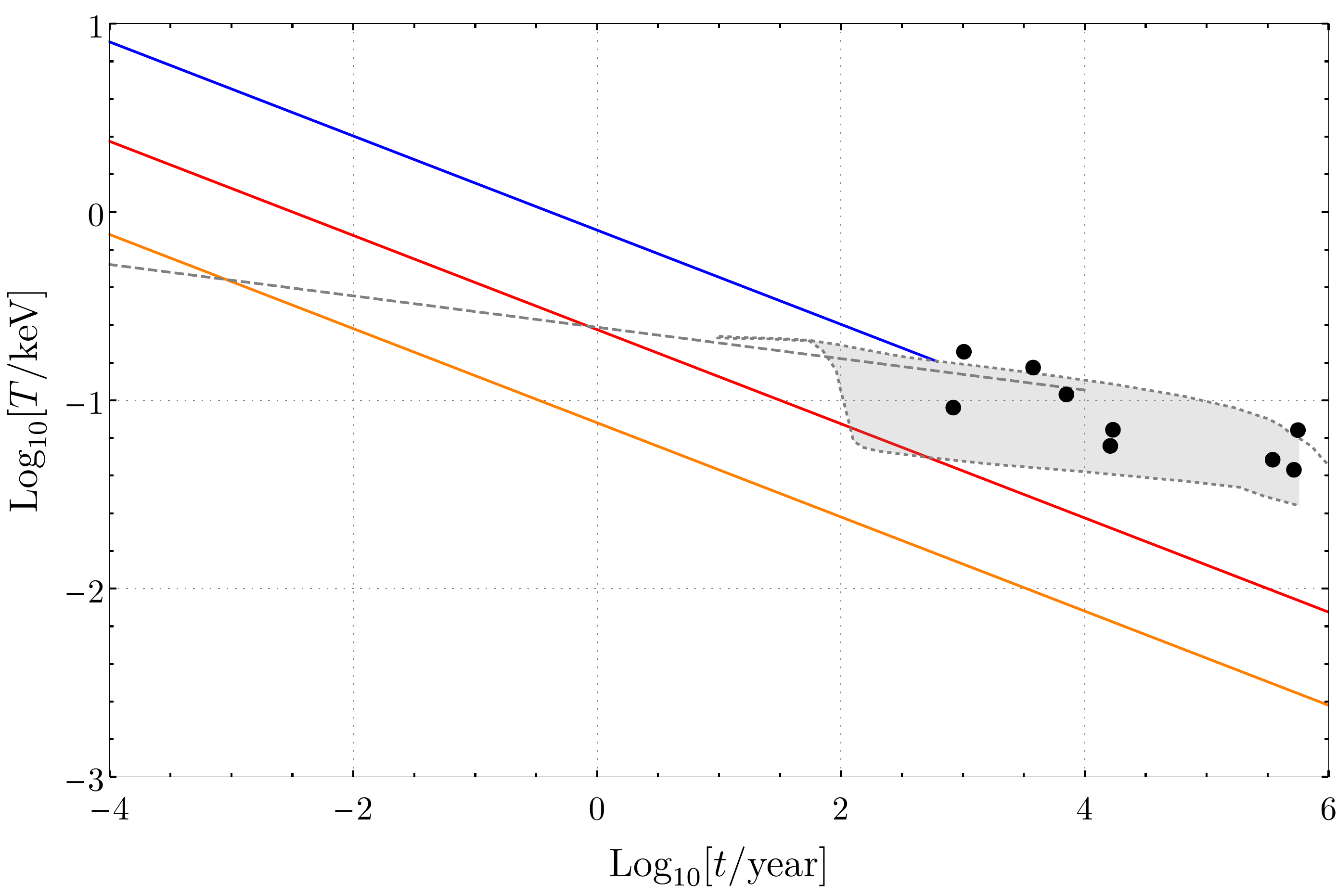}
\caption{Temperature of the gas inside bubbles of different maximal size as a function of time. The blue, red and orange solid lines correspond to bubbles with maximal size of 1, 10 and 100 light seconds, respectively. The evolution of the bubble is terminated when the bubble radius decreases to less than 200 km.  The grey dashed line shows the temperature of a neutron star cooling by emission of neutrinos in the neutron star core, while the gray shaded region shows the temperature of the neutron star during subsequent cooling of the neutron star by photon emission in different environments~\cite{Yakovlev:2004iq,Page:2004fy}. The black dots represent the temperature for some of the known neutron stars.}\label{fig:temp}
\end{figure}

During the bubble evolution, the thermal photons emitted far from the neutron star can have temperatures that differ from that of the neutron star (see figures~\ref{fig:temp}).  Therefore, it is possible to search for the existence of bubbles by measuring the thermal emission spectrum after a supernova explosion.
Since the shock wave passing through the bubble will tend to increase the size of the bubble, a smooth slowly contracting phase can only start after the shock wave has already passed through. 
After the shock wave and the surrounding matter have blown away, an observation of the final state neutron star and the bubble around it can be made~\footnote{It is not clear how long after a supernova one would have to wait.  For example, we still have not seen the neutron star/black hole created after supernova 1987A.}. In reality, we would likely have to wait a bit longer until when the emission of the shock wave can be spatially separated (if not spectroscopically separated) from that coming from the neutron star and the bubble since the emissions from the shock wave are likely very bright~\cite{Reynolds}.

Due to the large uncertainties in the initial velocity of the shock, we do not know the exact motion of the bubble after the shock has passed through it. However, this does not spoil our signature as the bubble will eventually enter a slowly contracting phase independent of its initial velocity.  During this phase, measurements of the luminosity as well as the frequency spectrum from the region around the neutron star when combined can tell us about the radius at which radiation is emitted.

\begin{figure}[t]
\centering
\includegraphics[scale=0.4]{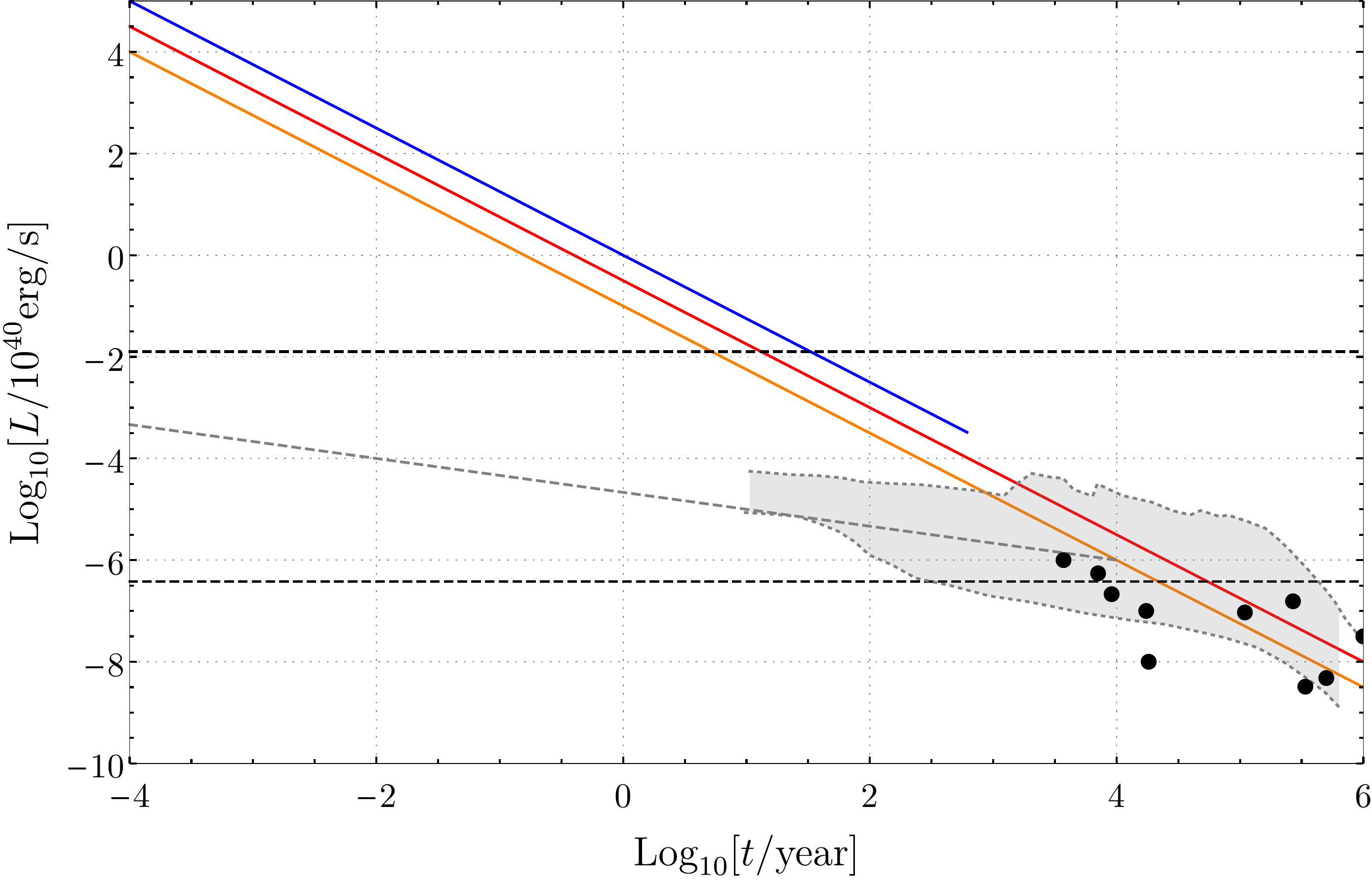}
\caption{Power emitted by the gas inside of bubbles of different maximal size as a function of time. The blue, red and orange solid lines correspond to bubbles with a maximal size of 1, 10 and 100 light seconds, respectively.  The grey dashed line shows the thermal power of a neutron star cooling dominantly by emission of neutrinos in the neutron star core, while the gray shaded region shows the power of thermal emission during subsequent cooling of the neutron star by photon emission in different environments~\cite{Yakovlev:2004iq,Page:2004fy}. The black dots represent the observed power for some of the known neutron stars. The black dashed lines shows the Eddington luminosity of a solar mass blackhole and the solar luminosity (top and bottom respectively).}\label{fig:power}
\end{figure}

If the existence of thermal radiation being emitted from a radius larger than the neutron star radius can be confirmed at a supernova remnant, young or old, follow-up studies are required to determine whether such a radius is the surface of a bubble or merely the exterior of some photon emission region.  If it were the surface of a bubble, then the bubble radius would decrease in a way that can be described by equation~\ref{eq:phase3evo}. In the beginning of this contraction phase, the temperature of the gas inside the bubble as well as the bubble radius should scale close to that of equation~\ref{eq:phase3scaling}.  As the gas and the neutron star continue to cool, the density profile of the gas inside the bubble should also start evolving. As the density profile deviates from a $1/R^3$ scaling, the temperature and density evolutions will also deviate from $T\propto t^{-1/4}$ and $R\propto t^{-1/8}$. A detailed simulation would be required to understand the exact evolution of the bubble radius and temperature.  However, independently from the exact bubble dynamics, during this last stage of the evolution, a $\delta m/m$ portion of the total progenitor mass will be released in X-rays over the whole period. A faster evolution~\cite{MacFadyen:1999mk} compared to that shown in figures~\ref{fig:temp} and~\ref{fig:power} or, in particular, a sudden change of the evolution process of the bubble would likely be an even more prominent a signal as a huge amount of power will likely be released from the supernova remnant in a very short amount of time. Moreover, independent of the exact form of the density profile, and correspondingly the temperature and luminosity as a function of time, we can use the measurement of the spectrum as a function of time to extract the most important information, the brane tension $\mathcal{T}$ and the energy difference $\Delta V$ with equation~\ref{eq:phase3evo}.

In the absence of a detailed simulation of the evolution of the density near the remnant neutron star, we will just show the temperature and luminosity as a function of time assuming a $1/R^3$ profile in figures ~\ref{fig:temp} and~\ref{fig:power} to demonstrate some of the important qualitative features for different size bubbles. These plots should not be taken as a prediction, especially beyond time scales that are close to the accretion time scale of a neutron star.

\subsection{Species dependent interactions}

In the previous sections, we have treated the bubble wall as a surface separating two regions where the mass of a particle differs by a fixed fractional amount $\delta m/m$ independent of whether it is a proton, a neutron or an electron. In reality, the matter coupling of the scalar $\phi$ likely breaks the equivalence principle and causes a species dependent mass shift. In this section, we will mainly discuss observable effects of two types of mass shifts, which can be looked for with spectral measurements of the neutron star.

The first generic type of couplings are those similar to the matter couplings of a QCD axion. These couplings modify the masses of the proton and neutron differently (see reference~\cite{Hook:2017psm} for more details) and affect minimally the properties of an electron. The mass differences between neutron and proton can change by a maximal amount of $10 \,{\rm MeV}$ if $\theta = \pi$. This will likely make several of the stable elements unstable inside the bubble surrounding the neutron star~\cite{1986AA}. The existence of oxygen, nitrogen, neon and, most importantly, iron can potentially be established near the surface of neutron stars through measurements of X-ray emission.  For example, if $\theta = \pi$ inside the bubble, ${}^{56}$Co will probably become lighter than ${}^{56}$Fe and the lifetime of iron would be short.  An observation of a ${\rm Fe\, K}\alpha$ line from regions inside the bubble would indicate the stability of iron and likely exclude this scenario.  Likewise, couplings that will make the neutron much lighter than the proton can also be constrained if abundant hydrogen lines are observed.

The second generic type of couplings are those that are similar to the matter coupling of a dilaton. These couplings result in a modification of the masses of all the standard model particles, and in particular the electron mass and therefore a fractional shift of all the characteristic emission lines in the form $\delta \omega/\omega = \delta m_e/m_e$ at leading order. Discovery of this type of overall shift is hard given that gravitational redshift has the same qualitative effect.  As a result, we can probably only look for couplings that modify the electron mass by close to unity, which cannot be faked by gravitational redshift around a neutron star with $GM_{\rm NS}/r_{\rm NS} \sim 0.1$.  Various other effects such as pressure broadening and Zeeman splittings must also similarly be well understood. Such a situation occurs when there are universal couplings of the scalar as well as for scalars with lepto-philic couplings, or couplings that modify the fine structure constants (see~\cite{Arvanitaki:2014faa} for more details on dilaton-like couplings to matter).

Localized regions where particle physics parameters differ from our own, or simply a spatially dependent fundamental constant have been looked for at great length in the past (see references in~\cite{Arvanitaki:2014faa}).  We suggest searching for spectroscopic features in localized regions with fundamental constants that differ from our own. To establish that a bubble is the reason behind these anomalous spectroscopic features, further investigation is required to discover the neutron star and the thermal emission properties from around the bubble wall.

\section{Phenomenology of non-spherical bubbles}\label{sec:collison}

In this section, we discuss deviations from the spherical approximation of isolated bubbles as well as bubble-bubble collisions and bubble-star collisions.  As answering most of the questions posed will require numerical simulations, we focus on describing qualitatively what occurs.

\subsection{Bubble non-sphericity} \label{sec:nonsphere}

\begin{figure}[ht]
\centering
\includegraphics[scale=0.6]{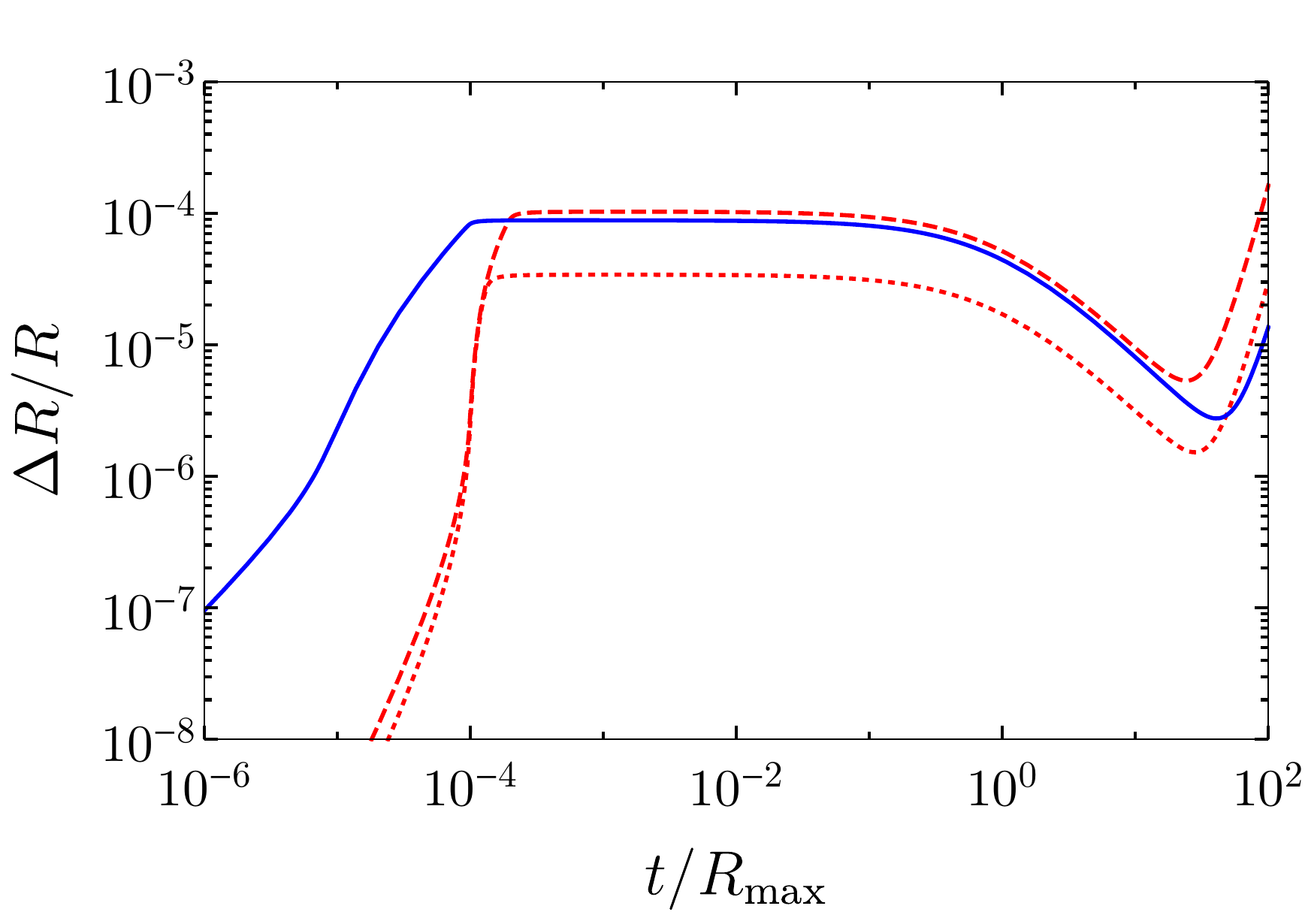}
\caption{We show how two bubbles with the same initial size can end up with much larger fractional differences in bubble radius ($\Delta R/R$).  The blue solid line shows how different in size two bubbles can get with different initial velocities. The two bubbles are assumed to start at $10\%$ and $90\%$ of the initial shock wave velocity. The red lines show how different in size two bubbles can get with different matter densities. The two bubbles are assumed to start in regions where the matter densities differ by $50\%$ (upper red dashed line) and $25\%$ (lower red dotted line), respectively.
}\label{fig:nonspherical}
\end{figure}

The growth of the bubble in a dense medium, unlike in empty space, can make a bubble very asymmetric. In the thin wall limit, the bubble evolution can be described by
modifying equation~\ref{eq:expansion} to a bubble that is not spherical symmetric. The non-sphericity of the bubbles in our case can arise due to two main reasons. Firstly, the distribution of the collapsing standard model matter near a supernova is likely not totally spherically symmetric, which will result in an asymmetric push on the bubble through matter interactions. Secondly, when the bubble speed is less than $(\delta m/m)^{1/2}$, the pressure difference between the two sides of the bubble, and therefore acceleration, grows with the velocity (see equation~\ref{eq:expansion}). Such a positive feedback might greatly enhance the non-sphericity. In figure \ref{fig:nonspherical}, we show the evolution of spherical bubbles which have slightly different initial velocity. These bubbles can grow to very different sizes at intermediate times. Though such an exercise does not take into account the effect of the brane tension, which favors a spherical bubble, it is likely that even if a spherical bubble is nucleated at rest, a significant asymmetry can arise as it accelerates into non-spherically symmetric matter. 

The bubble will become more and more spherically symmetric as it expands into regions with less matter density due to brane tension. However, the asymmetry can come back once the bubble starts to contract into matter, if the matter density is not fully spherically symmetric. Unlike the case of expansion where there is a positive feedback, during contraction, the friction force is proportional to the bubble velocity and therefore helps to smooth out the non-sphericity. Therefore, the bubble is very likely spherically symmetric when it falls back onto the neutron star.  A major complication is the effect of the passing shock wave, which, if non-spherical, would increase the non-sphericity as it passes through the bubble wall by $\mathcal{O}(1)$. A full 3D simulation is needed to understand how non-spherical the bubble is as the bubble expands and contracts, which is beyond the scope of this paper.

\subsection{Bubbles colliding with stars}

For the interactions we consider in this paper, a bubble never grows to be much larger than a few light hours, which is much smaller than the average distances between neutron stars and stars (neutron stars) in galaxies. Therefore, it is very unlikely that the bubble from a neutron star can collide with a nearby star or neutron star. In the rare case where a supernova happens in a binary, however, it might be possible for the bubble wall to interact with a star, a neutron star, or even another bubble. In this subsection, we will discuss some qualitative features of bubbles in a binary, focusing on bubble-star and bubble-bubble collisions. 

Bubbles colliding with stellar objects can occur when one of the stars in a binary undergoes a supernova. For example, supernovae can occur in binaries due to constant accretion from one star (usually a red giant) onto another (usually a white dwarf) and are thus a natural candidate. Though a typical type Ia supernovae usually does not lead to the formation of a neutron star, it is possible that supernovae can happen in a binary system (type Ib supernovae), or even a system with active accretion. In these types of systems, it is unclear how spherically symmetric the bubbles are, since the environment has matter densities that are very non-spherically symmetric.  Additionally, some of the assumptions, in particular the ones about the progenitor, used in section~\ref{sec:expansion} and~\ref{sec:contraction} may no longer be valid. In some parts of parameter space, the bubble can expand to be large enough that it collides with the companion.  Here, we will only discuss the case where the companion has properties similar to our Sun. The end result of the collision between the bubble and the star depends on the relative speed between the bubble and the companion.

Consider first the case where the bubble is traveling at non-relativistic speeds ($v_{\rm wall} \ll (\delta m/m)^{1/2}$).  As shown by equation~\ref{eq:expansion}, the reflection coefficient is large so that the majority of the star it encounters will be pushed away, while a small portion of the star will leak inside the bubble with $v \sim (\delta m/m)^{1/2}$.  The star will receive a kick from the collision with the bubble and some oscillatory modes of the star may be excited.  The part of the bubble wall that collides with the star will get accelerated due to this collision and the bubble will grow asymmetrically as a result.

If the bubble speed is ultra-relativistic ($\gamma_{\rm wall} \gg 1$), the bubble will instead swallow the majority of the star as it expands out. The star will be relatively unperturbed while the bubble itself will grow extremely asymmetric as the part of the bubble that encounters the star will receive a huge amount of energy from the $\delta m/m$ of the star it encounters. Such a bubble, as it contracts, will encounter the star again. Such a collision might lead to dramatic effects both to the star and also to the bubble itself. The bubble might even fragment into two bubbles during the collision as it grows to be extremely non-spherical.  We do not think analytical estimates can resolve what occurs in this situation.

One other case that might be constrained is the case where the bubble encounters the star when it has a speed of order $(\delta m/m)^{1/2}$ towards the end of its decelerating expansion phase. In this case, as the bubble enters the star, it might tear up the star as the transmission and reflection coefficients are both order unity if $(\delta m/m)^{1/2}$ is larger than the escape velocity of the star. The hydrodynamical stabilization of a star is likely completed destroyed even if $(\delta m/m)^{1/2}$ is slightly smaller than the escape velocity. It is unclear to us what will happen after the split as no example of this sort is known.

\subsection{Bubbles colliding with bubbles}

Bubble-bubble collisions can occur if supernova occurs when the progenitor is already in a binary with a neutron star. The expanding bubble from a supernova can potentially collide with the slowly contracting bubble surrounding the companion. A complete study of bubble-bubble collisions is beyond the scope of this paper. Here, we will only summarize some of the main qualitative features. As discussed in section~\ref{sec:nonsphere}, the bubble can grow to be very non-spherical as it expands, due to the density inhomogeneity during a supernova explosion. Therefore, as pointed out in~\cite{GarciaGarcia:2016xgv}, these bubble collisions can release a significant portion of its energy into gravitational waves.  Our discussion on bubble wall collisions will mirror bubble wall collisions that occur during phase transitions in the early universe (see ~\cite{Kleban:2011pg} for a recent review).

During a bubble collision, most of the energy released should go into a flux of scalars as the merging bubbles start to oscillate. The flux of scalars will be spread over a time scale that is at least the light crossing time of the bubble, that is, roughly the size of the binary system. In the case where the scalars are emitted with semi-relativistic speed from a collision, these scalars will be spread over a time scale that is the distance to the binary over the velocity dispersion. Either way, the flux of the emitted scalar will be quite independent of the micro-physics couplings that produce the bubble:

\begin{equation}
\mathcal{F} \sim \left(\frac{\delta m}{m}\right) \frac{M_{\rm progenitor}}{4 \pi t_{\rm flux} d_{\rm bi}^2},
\end{equation}
where $t_{\rm flux} > a_{\rm bi}$ is the time duration of the scalar flux, while $a_{\rm bi}$ and $d_{\rm bi}$ is the size of the binary and the distance from the binary to us, respectively. For a binary with $a_{\rm bi}$ of a few light seconds like that of a Hulse-Taylor pulsar binary~\cite{Taylor}, a binary that undergoes supernova explosion within $d_{\rm bi} \lesssim 10 \,{\rm kpc}$ could lead to fluxes comparable or larger than that of the dark matter on Earth with $\delta m / m \sim 10^{-2}$. This is very unlikely to happen since we have not seen a supernova explosion within a binary in our galaxy in the past thousands of years. Given that no DM experiments have been able to cut into the respective range of parameter space, we would have to be incredibly lucky to see such an event.

A subcomponent of the energy goes into gravitational waves. The emitted gravitational waves will be at a frequency that is roughly inverse the size of the binary. The amplitude of the gravitational wave from bubble collisions in a binary similar to that of the Hulse-Taylor binary will be at most of order
\begin{equation}
h \sim \left(\frac{\delta m}{m}\right) \frac{G M_{\rm progenitor}}{ d_{\rm bi}},
\end{equation}
if the bubble is $\mathcal{O}(1)$ asymmetric. For a binary that is $\mathcal{O}(10 \,{\rm kpc})$ away, the amplitude at Earth is at most $\mathcal{O}(10^{-19})$, with a frequency that can range from a few Hz to as small as $10^{-4}$ Hz. Such an amplitude might be observable as a transient by space-based gravitational wave detectors such as LISA or AGIS~\cite{Audley:2017drz,Dimopoulos:2008sv}. However, it is quite unlikely such an event takes place so close to us.  It might still be interesting to understand the collision of these bubbles, as future gravitational wave detectors might give us the chance to look for these collisions outside of our own galaxy.

\section{Conclusion and remarks}\label{sec:conclusion}

In this paper, we showed how a bubble containing a different minimum compared to the minimum we currently live in can be generated at the time of a supernova explosion, or similarly violent events where the density of matter undergoes a very sudden change. We described how such a bubble would evolve after a supernova explosion in the spherical bubble limit, and sketched how asymmetries can be generated for these confined bubbles. The bubble that gets generated during a core collapse supernova can accelerate to ultra-relativistic speed very quickly and reach sizes as large as $\sim 100 \,{\rm seconds}$ at the end of its expansion. The bubble eventually contracts into matter surrounding the neutron star and slowly loses all its energy through interactions with standard model matter. Such a process can last for thousands of years and leave striking signatures as a result of this slow release of energy into the environment.

The bubble can be potentially observable both right after a supernova explosion, as well as much later from its effect on the surrounding environment of the supernova remnant. Thermal and non-thermal radiation from a supernova can teach us a great deal about the potential existence of a confined bubble. We suggest measuring the emission spectrum of relatively young supernova remnants and how it evolves with time as a way to find evidence of a contracting bubble. However, these signatures depend strongly on how bubbles evolve, and details of the environment the progenitor is in, and detailed simulations are likely needed to compare with observation.

We have known for a long time that many stable and meta-stable vacua can exist, especially in various condensed matter systems. These vacua can appear and disappear depending on the environmental parameters of the system, leading to different forms of matter. Similarly, in quantum field theory, and especially in string theory, there can also be a wide variety of different vacua, whose properties can also depend strongly on the environment. The QCD axion and the Higgs are two prime examples where the potential changes dramatically as a result of the thermal effects in the early universe. 

In a recent paper~\cite{Hook:2017psm} and this paper, we discussed how matter densities can affect the minimum structure and the observational consequences of such a situation.  In a companion paper~\cite{NewMinimum2}, we will discuss how we can look for the existence of a new Higgs minimum with cosmological observations, providing exciting new probes into the nature of the Higgs potential at large field values.  The old examples and new possibilities we considered in the series of papers are summarized in the table~\ref{tab:vacua}. We hope many new exciting examples will emerge in the near future.
 
\begin{table}[ht]
\centering
\begin{tabular}{c|c|c}
 & Present & Past\\
New vacua appear & - & Axion potential turning on \& EWSB \\
New vacua disappear & Non-lethal Bubble & Higgs instability~\cite{Espinosa:2017sgp,NewMinimum2} \\
Vacua move (Potential flip) & NS Force~\cite{Hook:2017psm} & - \\
\end{tabular}
\caption{Scalar potentials can be affected by thermal and density corrections. Minima can appear as the universe cools down both during the electroweak symmetry breaking (EWSB) and QCD phase transition. Vacua can disappear due to both density and thermal effects and also move (flip sign) due to density effects.}\label{tab:vacua}
\end{table}

\section*{Acknowledgement}

The authors thank Matthew Johnson for many useful discussions as well as enormous help with simulation. The authors want to thank Masha Baryakhtar, Gustavo Marques-Tavares and Davide Racco for various clarifying discussions and comments on the draft. The authors would also like to thank Asimina Arvanitaki, David Curtin, William DeRocco, Savas Dimopoulos, Sergei Dubovsky, Jun Zhang for useful discussion. The authors want to acknowledge the Michigan LCTP, CERN for hospitality during the brainstorming stage of this project. JH wants to thank Maryland MCFP for hospitality during part of the project.   AH is supported in part by the NSF under Grant No. PHY-1620074 and by the Maryland Center for Fundamental Physics (MCFP).  Research at Perimeter Institute is supported
by the Government of Canada through Industry Canada
and by the Province of Ontario through the Ministry of
Economic Development \& Innovation.

\appendix
\section{Another explicit example of a landscape field}\label{sec:example}

In the main text, a model was presented where an axion was a {\it landscape field}.  In this Appendix, we demonstrate how a scalar modulating the mass of a fermion directly can also play the role of a landscape field.
For simplicity, we will take the fermion to be the electron, but it can be any fermion in the theory.  The model consists of $N$ copies of the Standard Model all exchanged via a $\mathbb{Z}_N$ symmetry while $\phi \rightarrow \phi + 2 \pi f/N$ under the same $\mathbb{Z}_N$ symmetry.
The landscape scalar, $\phi$, couples to the $N$ copies of the electron as
\bea
\mathcal{L} = \sum_{k=0}^{N-1} \left ( m_e - \epsilon \cos \left ( \frac{\phi}{f} + \frac{2 \pi k }{N} \right ) \right )  e_k e^c_k.
\eea
where $e^{(c)}_0$ is the electron and $e^{(c)}_j$ are the electrons in the other decoupled sectors.  The expectation value of $\phi$ couples to the electron mass so that the electron mass $m_e$ is different in the various minima of the theory fractionally by $\mathcal{O}(\epsilon/m_e)$.  Integrating out the fermions generates a potential for $\phi$
\bea
V_\phi \sim  m_e^4 \frac{\epsilon^N}{m_e^N} \cos \left ( \frac{N \phi}{f} \right ).
\eea
Additionally, we allow for a very soft breaking of the $\mathbb{Z}_N$ symmetry in the form of a potential term
$V_\phi = \delta^4 \cos \left ( \frac{\phi}{f} + \theta \right )$.  Thus, the final potential for the scalar $\phi$ is
\bea
\label{Eq: total}
V_\phi =   \frac{m_\phi^2 f^2}{N^2} \cos \left ( \frac{N \phi}{f} \right ) +  \delta^4 \cos \left ( \frac{\phi}{f} + \theta \right ).
\eea
where we are assuming that $m_\phi^2 f^2/N^2 \gg \delta^4$.

\paragraph{In empty space, there are many vacua that scan the cosmological constant}

The potential shown in Eq.~\ref{Eq: total} has a landscape that scans the CC with step sizes $\frac{\delta^4}{N}$.  Additionally, the vacua are all {\it ordered} so that the vacua with small CC are all close to each other in field space.

\paragraph{In empty space, tunneling between vacua is highly suppressed}

The vacuum tunneling rate depends exponentially sensitively on the bounce action
\bea
B \sim \frac{\Delta \phi^4}{\Delta V} \sim \frac{f^2}{N^2 m_\phi^2} \gg 1 .
\eea
In this paper, we consider regions of parameter space where $10^{15}$ GeV $\gtrsim f/N \gtrsim 10^{9}$ GeV and $m_\phi \lesssim 10^{-2}$ eV.  Clearly the tunneling rate is much longer than the age of the universe.

\paragraph{In medium, classical or quantum transitions between vacua are fast}

We first estimate the size of finite density effects on the potential of $\phi$.  Take the medium to have a number density of electrons $n_e$ and electrons to be non-relativistic.
\bea
\label{Eq: fdensity}
V _\text{finite density} = m_e (\phi) n_e =  \left ( m_e - \epsilon \cos \left ( \frac{\phi}{f} \right ) \right ) n_e
\eea

In order for the classical transition to be fast, we assume that the finite density effect is larger than the bare potential, $\epsilon n_e > m_\phi^2 f^2/N^2$.  If this is the case, then $\phi$ will classically roll to the bottom of the potential where $\phi = 0$.  We see that once all other parameters are fixed and as long as $m_\phi$ is small enough, finite density effects will always be significant and cause fast transitions.

\paragraph{Early universe constraints}

Obtaining the correct relic abundances of primordial atoms such as hydrogen and helium, places a constraint on any change of values of constants such as the electron mass.  Requiring that the in medium transitions between vacua are slow at Big Bang nucleosynthesis (BBN) places a weaker bound than requiring that the Sun does not source confined bubbles.  The reason is that the universe at BBN was less dense than the Sun.

The reheating temperature of these theories must be low.  At high temperatures, the finite density minima has a minimum at $\phi = 0$, see Eq.~\ref{Eq: fdensity}.  Therefore, in the early universe, $\phi \approx 0$.  Eventually the vacuum piece starts to dominate and $\phi$ oscillates around its minimum.  Estimating the energy density in $\phi$, one finds that it over-closes the universe.

\paragraph{Localization}

An important point about the finite density potential in Eq.~\ref{Eq: fdensity} is that its minima are misaligned with the true minimum outside of the finite density system.  Values were chosen so that at finite density, the true minimum is located at $\phi=0$.  However, in empty space, the true minimum is located at $\phi = - \theta f$.

In order to minimize the total energy, the field $\phi$ must transition to $\phi=0$ at large enough distances.  We have thus shown that the bubble will necessarily be localized around the object sourcing it.

\section{Fast oscillation and time reversal asymmetry} \label{Sec: fastosc}

In the main text, we described a slow contraction phase which ends with the bubble releasing a significant portion of its energy into heat.  
Depending on the details of the dynamics of the shock wave (see section~\ref{sec:shock}), the bubble might contract with enough kinetic energy to reach relativistic speed, and move through the standard model matter inside the bubble, and enter a period of fast oscillation.  The fast contraction phase is simply the time reversal of the fast expansion phase.
Similarly to the relativistic expansion phase, during the fast contraction phase the standard model matter the bubble encounters is mostly unperturbed.  What eventually stops the periods of fast expansion and contraction is the presence of friction.  A significant amount of friction can be generated during the non-relativistic part of the initial and final stage of the contraction/expansion phase.  This friction results from the large reflection coefficient present at low velocities.  As the bubble slows down, the matter that is reflected from the bubble wall does not encounter the bubble wall ever again, permanently removing energy from the system.

To estimate how much energy is lost per oscillation, we take the approximation that any particle reflected off the wall is assumed to never hit the wall again.  Thus, any energy that they extract from the wall is never re-injected into the system. There are other sources of friction present~\footnote{For example, particles with very small velocities cannot exit the wall because it doesn't satisfy energy conservation, while particles with very small velocities can enter the bubble. Similarly, particles that enter the bubble during expansion and exit during bubble contraction carry away energy from the bubble due to thermalization.}, but none remove parametrically more energy than this effect.
Consider a single particle at rest that encounters the bubble when it is expanding and has just become non-relativistic.  These particles remove an energy
\begin{eqnarray}
\delta E = \mathbb{R} \times \left(\frac{(2 k)^2}{2 m}\right) \approx \begin{cases} \frac{\delta m^2}{m v_{\rm wall}^2}, \quad & v_{\rm wall} \gtrsim (\delta m/m)^{1/2}, \\  m v_{\rm wall}^2, \quad & v_{\rm wall} \lesssim (\delta m/m)^{1/2}. \end{cases}
\end{eqnarray}
The majority of the energy will be removed when the wall has velocities $v_{\rm wall} \sim (\delta m/m)^{1/2}$ so that it is neither fully relativistic nor non-relativistic. This is why our equation~\ref{eq:expansion}, which describes the expansion and contraction in the non-relativistic and ultra-relativistic limit, has an apparent time reversal symmetry at leading order.

Combining the above result with the fact that the bubble expansion and contraction starts and finishes with a period of non-relativistic bubble motion, we find that the quality factor of the bubble evolution is 
\begin{equation}
Q \sim \frac{\delta m N_{\rm rel} }{ \delta E N_{\rm nr}} \lesssim \frac{m}{\delta m} \log \left[\frac{R_{\rm max}}{R_{\rm NS}}\right]
\end{equation}
for a $1/R^3$ density profile, where $N_{\rm nr}$ and  $N_{\rm rel}$ are the total number of particles the bubble passes through while non-relativistic and relativistic, respectively.

Because each oscillation lasts for only about $R_{\rm max}$ time, it is very unlikely that such a fast oscillating phase can last for long enough in the high-density region near the neutron star to be observable. Since the equilibrium radius between the brane tension and the pressure from the density of matter for a semi-relativistic bubble is very close to the maximum radius $R_{\rm max}$ of the bubble, the bubble will eventually end in the slow contracting phase starting from around $R_{\rm max}$ independently of the early stages of the evolution. 

\bibliographystyle{JHEP}
\bibliography{submit}

\providecommand{\href}[2]{#2}\begingroup\raggedright\begin{thebibliography}{10}

\bibitem{Weinberg:1988cp}
S.~Weinberg, {\it {The Cosmological Constant Problem}},  {\em Rev. Mod. Phys.}
  {\bf 61} (1989) 1--23.

\bibitem{Abbott:1984qf}
L.~F. Abbott, {\it {A Mechanism for Reducing the Value of the Cosmological
  Constant}},  {\em Phys. Lett.} {\bf 150B} (1985) 427--430.

\bibitem{Brown:1987dd}
J.~D. Brown and C.~Teitelboim, {\it {Dynamical Neutralization of the
  Cosmological Constant}},  {\em Phys. Lett.} {\bf B195} (1987) 177--182.

\bibitem{Steinhardt:2006bf}
P.~J. Steinhardt and N.~Turok, {\it {Why the cosmological constant is small and
  positive}},  {\em Science} {\bf 312} (2006) 1180--1182,
  [\href{http://arxiv.org/abs/astro-ph/0605173}{{\tt astro-ph/0605173}}].

\bibitem{Alberte:2016izw}
L.~Alberte, P.~Creminelli, A.~Khmelnitsky, D.~Pirtskhalava, and E.~Trincherini,
  {\it {Relaxing the Cosmological Constant: a Proof of Concept}},  {\em JHEP}
  {\bf 12} (2016) 022, [\href{http://arxiv.org/abs/1608.05715}{{\tt
  arXiv:1608.05715}}].

\bibitem{Graham:2019bfu}
P.~W. Graham, D.~E. Kaplan, and S.~Rajendran, {\it {Relaxation of the
  Cosmological Constant}},  \href{http://arxiv.org/abs/1902.06793}{{\tt
  arXiv:1902.06793}}.

\bibitem{Weinberg:1987dv}
S.~Weinberg, {\it {Anthropic Bound on the Cosmological Constant}},  {\em Phys.
  Rev. Lett.} {\bf 59} (1987) 2607.

\bibitem{Riess:1998cb}
{\bf Supernova Search Team} Collaboration, A.~G. Riess et~al., {\it
  {Observational evidence from supernovae for an accelerating universe and a
  cosmological constant}},  {\em Astron. J.} {\bf 116} (1998) 1009--1038,
  [\href{http://arxiv.org/abs/astro-ph/9805201}{{\tt astro-ph/9805201}}].

\bibitem{Perlmutter:1998np}
{\bf Supernova Cosmology Project} Collaboration, S.~Perlmutter et~al., {\it
  {Measurements of Omega and Lambda from 42 high redshift supernovae}},  {\em
  Astrophys. J.} {\bf 517} (1999) 565--586,
  [\href{http://arxiv.org/abs/astro-ph/9812133}{{\tt astro-ph/9812133}}].

\bibitem{Spergel:2003cb}
{\bf WMAP} Collaboration, D.~N. Spergel et~al., {\it {First year Wilkinson
  Microwave Anisotropy Probe (WMAP) observations: Determination of cosmological
  parameters}},  {\em Astrophys. J. Suppl.} {\bf 148} (2003) 175--194,
  [\href{http://arxiv.org/abs/astro-ph/0302209}{{\tt astro-ph/0302209}}].

\bibitem{Bousso:2007gp}
R.~Bousso, {\it {TASI Lectures on the Cosmological Constant}},  {\em Gen. Rel.
  Grav.} {\bf 40} (2008) 607--637, [\href{http://arxiv.org/abs/0708.4231}{{\tt
  arXiv:0708.4231}}].

\bibitem{Coleman:1980aw}
S.~R. Coleman and F.~De~Luccia, {\it {Gravitational Effects on and of Vacuum
  Decay}},  {\em Phys. Rev.} {\bf D21} (1980) 3305.

\bibitem{Coleman:1977py}
S.~R. Coleman, {\it {The Fate of the False Vacuum. 1. Semiclassical Theory}},
  {\em Phys. Rev.} {\bf D15} (1977) 2929--2936. [Erratum: Phys.
  Rev.D16,1248(1977)].

\bibitem{Callan:1977pt}
C.~G. Callan, Jr. and S.~R. Coleman, {\it {The Fate of the False Vacuum. 2.
  First Quantum Corrections}},  {\em Phys. Rev.} {\bf D16} (1977) 1762--1768.

\bibitem{Bousso:2000xa}
R.~Bousso and J.~Polchinski, {\it {Quantization of four form fluxes and
  dynamical neutralization of the cosmological constant}},  {\em JHEP} {\bf 06}
  (2000) 006, [\href{http://arxiv.org/abs/hep-th/0004134}{{\tt
  hep-th/0004134}}].

\bibitem{Polchinski:2006gy}
J.~Polchinski, {\it {The Cosmological Constant and the String Landscape}},  in
  {\em {The Quantum Structure of Space and Time: Proceedings of the 23rd Solvay
  Conference on Physics. Brussels, Belgium. 1 - 3 December 2005}},
  pp.~216--236, 2006.
\newblock \href{http://arxiv.org/abs/hep-th/0603249}{{\tt hep-th/0603249}}.

\bibitem{Csaki:2018fls}
C.~Csáki, C.~Eröncel, J.~Hubisz, G.~Rigo, and J.~Terning, {\it {Neutron Star
  Mergers Chirp About Vacuum Energy}},
  \href{http://arxiv.org/abs/1802.04813}{{\tt arXiv:1802.04813}}.

\bibitem{Hook:2017psm}
A.~Hook and J.~Huang, {\it {Probing axions with neutron star inspirals and
  other stellar processes}},  {\em JHEP} {\bf 06} (2018) 036,
  [\href{http://arxiv.org/abs/1708.08464}{{\tt arXiv:1708.08464}}].

\bibitem{Graham:2015cka}
P.~W. Graham, D.~E. Kaplan, and S.~Rajendran, {\it {Cosmological Relaxation of
  the Electroweak Scale}},  {\em Phys. Rev. Lett.} {\bf 115} (2015), no.~22
  221801, [\href{http://arxiv.org/abs/1504.07551}{{\tt arXiv:1504.07551}}].

\bibitem{Arvanitaki:2016xds}
A.~Arvanitaki, S.~Dimopoulos, V.~Gorbenko, J.~Huang, and K.~Tilburg, {\it {A
  small weak scale from a small cosmological constant}},  {\em JHEP} {\bf 05}
  (2017) 071, [\href{http://arxiv.org/abs/1609.06320}{{\tt arXiv:1609.06320}}].

\bibitem{Hook:2018jle}
A.~Hook, {\it {Solving the Hierarchy Problem Discretely}},  {\em Phys. Rev.
  Lett.} {\bf 120} (2018), no.~26 261802,
  [\href{http://arxiv.org/abs/1802.10093}{{\tt arXiv:1802.10093}}].

\bibitem{Arvanitaki:2009fg}
A.~Arvanitaki, S.~Dimopoulos, S.~Dubovsky, N.~Kaloper, and J.~March-Russell,
  {\it {String Axiverse}},  {\em Phys. Rev.} {\bf D81} (2010) 123530,
  [\href{http://arxiv.org/abs/0905.4720}{{\tt arXiv:0905.4720}}].

\bibitem{Hu:2000ke}
W.~Hu, R.~Barkana, and A.~Gruzinov, {\it {Cold and fuzzy dark matter}},  {\em
  Phys. Rev. Lett.} {\bf 85} (2000) 1158--1161,
  [\href{http://arxiv.org/abs/astro-ph/0003365}{{\tt astro-ph/0003365}}].

\bibitem{Hui:2016ltb}
L.~Hui, J.~P. Ostriker, S.~Tremaine, and E.~Witten, {\it {Ultralight scalars as
  cosmological dark matter}},  {\em Phys. Rev.} {\bf D95} (2017), no.~4 043541,
  [\href{http://arxiv.org/abs/1610.08297}{{\tt arXiv:1610.08297}}].

\bibitem{FuzzyKen}
A.~Arvanitaki, S.~Dimopoulos, M.~Galanis, L.~Lehner, J.~Thompson, and
  K.~Van~Tilburg, {\it {in preparation}}, .

\bibitem{Preskill:1982cy}
J.~Preskill, M.~B. Wise, and F.~Wilczek, {\it {Cosmology of the Invisible
  Axion}},  {\em Phys. Lett.} {\bf B120} (1983) 127--132. [,URL(1982)].

\bibitem{Espinosa:2007qp}
J.~R. Espinosa, G.~F. Giudice, and A.~Riotto, {\it {Cosmological implications
  of the Higgs mass measurement}},  {\em JCAP} {\bf 0805} (2008) 002,
  [\href{http://arxiv.org/abs/0710.2484}{{\tt arXiv:0710.2484}}].

\bibitem{Espinosa:2015qea}
J.~R. Espinosa, G.~F. Giudice, E.~Morgante, A.~Riotto, L.~Senatore, A.~Strumia,
  and N.~Tetradis, {\it {The cosmological Higgstory of the vacuum
  instability}},  {\em JHEP} {\bf 09} (2015) 174,
  [\href{http://arxiv.org/abs/1505.04825}{{\tt arXiv:1505.04825}}].

\bibitem{Franciolini:2018ebs}
G.~Franciolini, G.~F. Giudice, D.~Racco, and A.~Riotto, {\it {Implications of
  the detection of primordial gravitational waves for the Standard Model}},
  \href{http://arxiv.org/abs/1811.08118}{{\tt arXiv:1811.08118}}.

\bibitem{NewMinimum2}
A.~Hook, J.~Huang, and D.~Racco, {\it {in preparation}}, .

\bibitem{Huang:2018pbu}
J.~Huang, M.~C. Johnson, L.~Sagunski, M.~Sakellariadou, and J.~Zhang, {\it
  {Prospects for axion searches with Advanced LIGO through binary mergers}},
  \href{http://arxiv.org/abs/1807.02133}{{\tt arXiv:1807.02133}}.

\bibitem{Kamionkowski:1992mf}
M.~Kamionkowski and J.~March-Russell, {\it {Planck scale physics and the
  Peccei-Quinn mechanism}},  {\em Phys. Lett.} {\bf B282} (1992) 137--141,
  [\href{http://arxiv.org/abs/hep-th/9202003}{{\tt hep-th/9202003}}].

\bibitem{Alarcon:2011zs}
J.~M. Alarcon, J.~Martin~Camalich, and J.~A. Oller, {\it {The chiral
  representation of the $\pi N$ scattering amplitude and the pion-nucleon sigma
  term}},  {\em Phys. Rev.} {\bf D85} (2012) 051503,
  [\href{http://arxiv.org/abs/1110.3797}{{\tt arXiv:1110.3797}}].

\bibitem{Linde:1981zj}
A.~D. Linde, {\it {Decay of the False Vacuum at Finite Temperature}},  {\em
  Nucl. Phys.} {\bf B216} (1983) 421. [Erratum: Nucl. Phys.B223,544(1983)].

\bibitem{Linde:1980tt}
A.~D. Linde, {\it {Fate of the False Vacuum at Finite Temperature: Theory and
  Applications}},  {\em Phys. Lett.} {\bf 100B} (1981) 37--40.

\bibitem{Dine:1992wr}
M.~Dine, R.~G. Leigh, P.~Y. Huet, A.~D. Linde, and D.~A. Linde, {\it {Towards
  the theory of the electroweak phase transition}},  {\em Phys. Rev.} {\bf D46}
  (1992) 550--571, [\href{http://arxiv.org/abs/hep-ph/9203203}{{\tt
  hep-ph/9203203}}].

\bibitem{Chang:2018rso}
J.~H. Chang, R.~Essig, and S.~D. McDermott, {\it {Supernova 1987A Constraints
  on Sub-GeV Dark Sectors, Millicharged Particles, the QCD Axion, and an
  Axion-like Particle}},  {\em JHEP} {\bf 09} (2018) 051,
  [\href{http://arxiv.org/abs/1803.00993}{{\tt arXiv:1803.00993}}].

\bibitem{Arvanitaki:2014faa}
A.~Arvanitaki, J.~Huang, and K.~Van~Tilburg, {\it {Searching for dilaton dark
  matter with atomic clocks}},  {\em Phys. Rev.} {\bf D91} (2015), no.~1
  015015, [\href{http://arxiv.org/abs/1405.2925}{{\tt arXiv:1405.2925}}].

\bibitem{1986AA}
F.-K. {Thielemann}, K.~{Nomoto}, and K.~{Yokoi}, {\it {Explosive
  nucleosynthesis in carbon deflagration models of Type I supernovae}},  {\em
  Astronomy \& Astrophysics} {\bf 158} (Apr., 1986) 17--33.

\bibitem{RevModPhys.62.801}
H.~A. Bethe, {\it Supernova mechanisms},  {\em Rev. Mod. Phys.} {\bf 62} (Oct,
  1990) 801--866.

\bibitem{1976ApJS}
T.~A. {Weaver}, {\it {The structure of supernova shock waves}},  {\em The
  Astrophysical Journal, Supplement} {\bf 32} (Oct., 1976) 233--282.

\bibitem{2015MNRAS.453..287M}
B.~{M{\"u}ller}, {\it {The dynamics of neutrino-driven supernova explosions
  after shock revival in 2D and 3D}},  {\em Monthly Notices of the Royal
  Astronomical Society} {\bf 453} (Oct, 2015) 287--310,
  [\href{http://arxiv.org/abs/1506.05139}{{\tt arXiv:1506.05139}}].

\bibitem{Yakovlev:2004iq}
D.~G. Yakovlev and C.~J. Pethick, {\it {Neutron star cooling}},  {\em Ann. Rev.
  Astron. Astrophys.} {\bf 42} (2004) 169--210,
  [\href{http://arxiv.org/abs/astro-ph/0402143}{{\tt astro-ph/0402143}}].

\bibitem{Reynolds}
S.~P. Reynolds, {\it Supernova remnants at high energy},  {\em Annual Review of
  Astronomy and Astrophysics} {\bf 46} (2008), no.~1 89--126,
  [\href{http://arxiv.org/abs/https://doi.org/10.1146/annurev.astro.46.060407.145237}{{\tt
  https://doi.org/10.1146/annurev.astro.46.060407.145237}}].

\bibitem{Page:2004fy}
D.~Page, J.~M. Lattimer, M.~Prakash, and A.~W. Steiner, {\it {Minimal cooling
  of neutron stars: A New paradigm}},  {\em Astrophys. J. Suppl.} {\bf 155}
  (2004) 623--650, [\href{http://arxiv.org/abs/astro-ph/0403657}{{\tt
  astro-ph/0403657}}].

\bibitem{MacFadyen:1999mk}
A.~I. MacFadyen, S.~E. Woosley, and A.~Heger, {\it {Supernovae, jets, and
  collapsars}},  {\em Astrophys. J.} {\bf 550} (2001) 410,
  [\href{http://arxiv.org/abs/astro-ph/9910034}{{\tt astro-ph/9910034}}].

\bibitem{GarciaGarcia:2016xgv}
I.~Garcia~Garcia, S.~Krippendorf, and J.~March-Russell, {\it {The String
  Soundscape at Gravitational Wave Detectors}},  {\em Phys. Lett.} {\bf B779}
  (2018) 348--352, [\href{http://arxiv.org/abs/1607.06813}{{\tt
  arXiv:1607.06813}}].

\bibitem{Kleban:2011pg}
M.~Kleban, {\it {Cosmic Bubble Collisions}},  {\em Class. Quant. Grav.} {\bf
  28} (2011) 204008, [\href{http://arxiv.org/abs/1107.2593}{{\tt
  arXiv:1107.2593}}].

\bibitem{Taylor}
J.~H. {Taylor} and J.~M. {Weisberg}, {\it {A new test of general relativity -
  Gravitational radiation and the binary pulsar PSR 1913+16}},  {\em The
  Astrophysical Journal} {\bf 253} (Feb., 1982) 908--920.

\bibitem{Audley:2017drz}
{\bf LISA} Collaboration, H.~Audley et~al., {\it {Laser Interferometer Space
  Antenna}},  \href{http://arxiv.org/abs/1702.00786}{{\tt arXiv:1702.00786}}.

\bibitem{Dimopoulos:2008sv}
S.~Dimopoulos, P.~W. Graham, J.~M. Hogan, M.~A. Kasevich, and S.~Rajendran,
  {\it {An Atomic Gravitational Wave Interferometric Sensor (AGIS)}},  {\em
  Phys. Rev.} {\bf D78} (2008) 122002,
  [\href{http://arxiv.org/abs/0806.2125}{{\tt arXiv:0806.2125}}].

\bibitem{Espinosa:2017sgp}
J.~R. Espinosa, D.~Racco, and A.~Riotto, {\it {Cosmological Signature of the
  Standard Model Higgs Vacuum Instability: Primordial Black Holes as Dark
  Matter}},  {\em Phys. Rev. Lett.} {\bf 120} (2018), no.~12 121301,
  [\href{http://arxiv.org/abs/1710.11196}{{\tt arXiv:1710.11196}}].

\end{thebibliography}\endgroup

\end{document}